\documentclass[12pt,draftclsnofoot,onecolumn]{IEEEtran}
\usepackage[utf8]{inputenc}
\usepackage{cite}
\usepackage{floatrow}
\usepackage{amsmath,amssymb,amsfonts,stfloats}
\usepackage{tabularx}
\usepackage{algorithmic}
\usepackage{graphicx}
\usepackage{textcomp}
\usepackage{floatrow}
\usepackage{xcolor}
\usepackage{booktabs}

\usepackage{multicol}
\usepackage{acronym}
\usepackage{url}
\usepackage{mathtools}
\usepackage[caption=false,font=footnotesize]{subfig}

\def\BibTeX{{\rm B\kern-.05em{\sc i\kern-.025em b}\kern-.08em
    T\kern-.1667em\lower.7ex\hbox{E}\kern-.125emX}}

\begin{document}
 
%\title{Experimental Demonstration of a Dual Domain Waveform Design for Integrated Sensing and Communication Systems}
\title{Integrated Sensing and Communication System \\via Dual-Domain Waveform Superposition }

\author{\IEEEauthorblockN{Dario Tagliaferri, Marouan Mizmizi,  Silvia Mura, Francesco Linsalata, Davide Scazzoli, Damiano Badini, Maurizio Magarini and Umberto Spagnolini}}
%\IEEEauthorblockA{Dipartimento di Elettronica, Informazione e Bioingegneria, Politecnico di Milano, Via Ponzio 34/5, 20133, Milano Italy}
%E-mails: \{luca.rinaldi,dario.tagliaferri,francesco.linsalata,marouan.mizmizi,maurizio.magarini,umberto.spagnolini\}@polimi.it}

\maketitle

\begin{abstract}

Integrated Sensing and Communication (ISAC) systems are recognised as one of the key ingredients of the sixth generation (6G) network. A challenging topic in ISAC is the design of a single waveform combining both communication and sensing functionalities on the same time-frequency-space resources, allowing to tune the performance of both with partial or full hardware sharing.    
This paper proposes a dual-domain waveform design approach that superposes onto the frequency-time (FT) domain both the legacy orthogonal frequency division multiplexing (OFDM) signal and a sensing one, purposely designed in the delay-Doppler domain. With a proper power downscaling of the sensing signal w.r.t. OFDM, it is possible to exceed regulatory bandwidth limitations proper of legacy multicarrier systems to increase the sensing performance while leaving communication substantially unaffected. Numerical and experimental results prove the effectiveness of the dual-domain waveform, notwithstanding a power abatement of at least 30 dB of the signal used for sensing compared to the one used for  communication. The dual-domain ISAC waveform outperforms both OFDM and orthogonal time-frequency-space (OTFS) in terms of Cram\'{e}r-Rao bound on delay estimation (up to 20 dB), thanks to its superior resolution, with a negligible penalty on the achievable rate.

\end{abstract}

%\begin{IEEEkeywords}
%Integrated sensing and communication, 6G, waveform design, experimental testbed
%\end{IEEEkeywords}

\maketitle

\section{Introduction}\label{sect:introduction}

%\textcolor{red}{lo sto rileggendo tutto, penso di finire stasera e al piu domani faccio abstract\\ commento: faccio fatica a determinare cosa davvero stiamo dicendo coi risultati, cioe' io lo so e l ho capito ma leggendolo non riesco a vederlo senza leggere il commento finale, per me va snellita quella parte. \\
%commento: la parte di allocazione di potenza ora mi piace molto invece, non la modificherei piu\\ }
%ISAC con waveform perchè si fa \\
%cosa facciamo noi: cosa è dual domain e superposition \\
%ottimizzazione \\
%risultati e vantaggi rispetto a OFDM e OTFS con numeri alla mano \\
%funziona anche HW

While the fifth generation (5G) network rollout is in full swing, academia and industry have already initiated to speculate on the next generation of wireless communication, namely 6G \cite{Saad2020AVO}. Following the general trend of successive generations of communication systems, 6G is expected to introduce new services, e.g., extended reality, high-fidelity hologram, and digital twin, with more stringent requirements. Current spectrum allocations will run out swiftly once 6G use cases start to be developed, which open up to millimeter-wave (mmWave, $30-300$ GHz) and sub-THz ($> 100$ GHz) bands to enable terabit connections. 

Pursuing higher frequencies and bandwidths paves the way to high-resolution sensing, and 6G is expected to be the first generation of wireless networks with sensing embedded as a service~\cite{Prelcic2020leveraging, Heath2021, wild2021JCS6G, Zhang2022EnablingJC}. The research on integrated sensing and communication (ISAC) systems has accelerated in the last decade due to a renewed interest in fusing the communication and sensing functionalities into a single waveform, with complete sharing of frequency/time/space and hardware resources~\cite{Heath2021overview}. The most general approach to ISAC is optimizing one functionality leaving the other constrained by a suitable metric, with an inherent trade-off between communication and sensing performance to be tuned. Nevertheless, for 6G networks, communication is the primary functionality, while sensing is added on top and shall not interfere. This approaach is known as communication-centric ISAC and is the focus of this paper \cite{Liu_survey}.

\subsection{Related works}

In communication-centric ISAC systems, the design of a single waveform is currently one of the most challenging research aspects~\cite{Liu_survey}. The goal is to add sensing functionality on top of the communication waveform, e.g., orthogonal frequency division multiplexing (OFDM), possibly with modifications, to support beam and blockage management (reduction of the beam training time, and proactive blockage prediction).
Waveform design via beampattern optimization refers to ISAC systems in which the spatial correlation of the signal across the transmitting (Tx) antennas is designed to guarantee a flexible beampattern for both communication and sensing purposes \cite{liu2018toward,Barneto2020_bf_and_waveform,Puccietal2022}. 
%For example, the works in \cite{Barneto2020_bf_and_waveform,barneto2020multibeam, Pucci:21,Puccietal2022} propose a multibeam ISAC system where beamformers are optimized via proper power splitting to simultaneously allow a steady communication beam towards the users and a sensing beam to scan the environment. The authors of ~\cite{liu2018beampattern} optimize the communication beamformer such that the achieved beam pattern matches the radar’s while satisfying the communication performance requirements.  
%An extension to analog arrays can be found in~\cite{zhang2018multibeam} \textcolor{red}{la parte spaziale non e' trattata cosi in dettaglio da metterci 4 refs nello soa secondo me}. 
Another approach consists of directly optimizing the resource allocation in the frequency-time (FT) domain, by employing multicarrier signals, such as OFDM, whose usage for sensing is explored in \cite{Sturm2011,Puccietal2022,Pucci:21,BarnetoFullDuplex}. The seminal work in \cite{Sturm2011} was the first to suggest a signal processing algorithm for an OFDM-based radar. The work in \cite{Shi2018_powermin_OFDM_coexistence} proposes three power minimization-based OFDM radar waveform designs for the coexistence between different radar and communication terminals on the same spectrum. The authors of \cite{Yongjun2017_OFDMdesign_MI} employ information-theoretic metrics applied to the communication and the sensing channels to design the OFDM ISAC waveform. Differently, works \cite{Bica2015_opportunistic,Bica2019_multicarrier} consider splitting the available OFDM subcarriers into radar and communication subsets, optimizing the radar mutual information according to a communication or sensing-centric policy. The trade-off between the two functionalities is that the radar subcarriers are used for sensing and the others for communication. 
Works \cite{Puccietal2022,Pucci:21} analyze the ISAC performance capabilities of 5G OFDM waveform, considering fully digital arrays and multi-beam design to split the spatial resources between communication and sensing. In \cite{BarnetoFullDuplex}, the authors discuss the practical issues of 5G OFDM for delay/Doppler estimation, e.g., complexity of the processing chain, self-interference, etc. They demonstrate, with experimental measurements, the feasibility of OFDM range/Doppler imaging. 
A leap forward has been made in~\cite{Barneto2021_Optimized_Wave}. where the authors propose to fill the empty communication subcarriers with sensing pilots (i.e., radar subcarriers). The power and phase of radar subcarriers are optimized by minimizing the Cram\'{e}r-Rao bound (CRB) on the delay and Doppler estimation for a single target while limiting the peak-to-average power ratio. 
Another relevant contribution in waveform design using optimized OFDM is in \cite{Wymeersch2021}. Here, the delay and Doppler CRB for multiple targets are derived, with optimal time/frequency/power resource allocation. The authors point out the problem of the ambiguity function of the generated waveform, whose sidelobes could mask weak targets that are not known during optimization. 

More recently, orthogonal time-frequency-space (OTFS) has been considered as a performing candidate waveform for sensing purposes on top of communication ones \cite{Raviteja_OTFS_radar}. Indeed, thanks to its structure it is possible to easily localize the targets in the delay-Doppler (DD) domain, providing sensing performance comparable to MIMO radars. In contrast to OFDM, in OTFS modulation data symbols are placed in the DD domain, overcoming the OFDM issues in doubly-selective channels~\cite{Saif2021_OTFS}. In OTFS, Doppler is exploited as an additional source of diversity, reducing pilot overhead without introducing inter-symbol interference. However, substantial modification of the current 5G new radio (NR) is necessary to introduce OTFS, requiring processing bursts of consecutive OFDM symbols, and this is in contrast with the low latency of many 6G services.
In \cite{gaudio2020_OFDM_vs_OTFS}, the authors derived an approximated maximum likelihood algorithm and the corresponding CRB for an OTFS ISAC system. The work in \cite{otfsISAC2022} proposes an optimized transmission framework based on the spatially spread OTFS modulation, leveraging the inherent difference between communication and sensing channels (the path
with the strongest echo power for radar sensing may not be the strongest path for communication).

%The results show that OTFS appears to be particularly suited for achieving high information rates and near-optimal sensing estimation performance. 
%Motivated by the mismatch of the reflection strengths between the radar sensing and communication (the path with the strongest echo power for radar sensing may not be the strongest path for communication) the recent work in \cite{otfsISAC2022} designs and optimizes a transmission framework based on the spatially spread OTFS modulation. 

In all the aforementioned research works, the sensing accuracy on delay/range and Doppler/velocity estimation is known to be limited by the allocated bandwidth and signal burst duration respectively. The former is set by regulatory limits \cite{ETSI_BSrequirements}. For instance, the OFDM-based 5G NR standard at mmWave (28 GHz) employs contiguous spectra of $400$ MHz (up to $1600$ MHz with carrier aggregation), although the potentially available bandwidth is $3$ GHz \cite{ETSIphy}. Moreover, in practical systems the Rx processing is based on discrete Fourier transform (DFT), thus the CRBs are rarely achieved. In these settings, the CRB is not the only figure of metric to be considered and the ambiguity function determines instead the practical resolution capabilities. In addition, allocating OFDM resources in FT with unequal power, possibly with an occupation factor $<100\%$, leads to high sidelobes in the ambiguity function, with a non-negligible amount of energy spread outside the main lobe. The OFDM waveform design by constraining the integrated sidelobe level of the ambiguity function, as in \cite{Wymeersch2021}, leads to complex optimizations increasing the computational burden, calling for simple yet effective approaches.   

%\textcolor{red}{Maximum single bandwdith is 200MHz for FR1 and 400MHz for FR2, with a maximum of 1600MHz with carrier aggregation [TS 38.101] - Metterei la reference}
%Contributi 

\subsection{Contributions}

In this paper, we propose a novel waveform for monostatic ISAC setups at the base station (BS). The waveform is obtained by superposition, in the FT domain, of the conventional OFDM signal, used for communication, and a dedicated sensing signal. The latter consists of a single pulse in the DD domain that maps into a 2D sinusoid in the FT domain. This limits the latency and minimises the computational complexity. The sensing signal is power-optimized in order to not interfere with legacy OFDM at the user equipment (UE) side. The advantages of the proposed \textit{dual-domain} approach are threefold: \textit{(i)} with a suitable power allocation of the communication and (mostly) sensing signals, it is possible to exceed regulatory bandwidth limits improving the delay/range resolution of the ISAC system, leaving the achievable rate almost unaffected; \textit{(ii)} employing the 2D sinusoid for sensing, the sidelobes of the ambiguity function are minimal, thanks to the whole coverage of FT resources (differently from legacy OFDM); \textit{(iii)} the proposed method does not require additional processing at the UE side and it requires only an IDFT-DFT pair at the BS (sensing Rx) side. 

%Notice that the DD domain is the basis of orthogonal time frequency space (OTFS) modulation, that has been proven to outperform OFDM in double-selective channels~\cite{Saif2021_OTFS} and to provide comparable sensing performance to MIMO radars~\cite{gaudio2019performance,Yuan2021_OTFS_vehicular}. However, placing information symbols in DD domain as in OTFS requires processing bursts of data, increasing the latency at Rx as well as its computational burden~\cite{Raviteja_OTFS_data_decoding,gaudio2020_OFDM_vs_OTFS}. Herein, the DD domain is not used for information symbols, but rather for the sensing signal only. 

%\textcolor{red}{Forse un paio di ref nella parte related works sulla OTFS le metterei...}

The novel contributions and results of the paper can be summarized as follows:
\begin{itemize}

    \item We propose and design an ISAC waveform combining the legacy OFDM signal with a dedicated sensing signal properly power-scaled and superposed in the FT domain. With a suitable power allocation (at least 30 dB less compared to OFDM), the sensing signal can exceed regulatory spectrum constraints to increase the range resolution of the ISAC system, fulfilling adjacent channel leakage ratio (ACLR) constraints \cite{ETSI_BSrequirements}. We formulate a convex power allocation problem under sensing and communication constraints, which is easily solvable in practical systems.
    
    \item We compare the performance of the dual-domain, OFDM and OTFS ISAC waveforms in terms of CRB on delay and Doppler estimation, mainlobe-to-total-energy ratio (MTER) of the ambiguity function, and achievable rate. The MTER measures both the improved resolution brought by the proposed waveform and the amount of energy in the sidelobes. The results highlight that the proposed dual-domain ISAC waveform can be effectively used in the near-to-medium range (up to $60-80$ m range @$30$ GHz carrier frequency). Moreover, when the sensing signal occupies 5x (or more) bandwidth compared to OFDM/OTFS, the dual-domain ISAC system exhibits negligible penalty in both CRB (computed under a single target assumption) and achievable rate compared to OFDM and OTFS, increasing the range resolution up to a factor 5. Moreover, the performance in the case of two targets reveals that, in some conditions, the CRB on delay estimation of the dual-domain waveform is even lower than that of OFDM/OTFS.  
    
    \item We demonstrate the capabilities of the proposed ISAC system with a dedicated experimental test using off-the-shelf mm-Wave communication transceivers, showing the promising application to practical systems.

\end{itemize}

\subsection{Organization}
The remainder of the paper is organized as follows: Section \ref{sect:system_model} outlines the considered scenario and the channel model, Section \ref{sect:TxISAC} describes the proposed Tx ISAC waveform, Section \ref{sect:Rx_signal} details the Rx communication and sensing signals at BS and UE, Section \ref{sect:power_opt} outlines the optimal ISAC power allocation, Section \ref{sect:numerical_results} reports and discusses the simulation results while Section \ref{sect:experimental_results} shows the experimental demonstration. Finally, Section \ref{sect:conclusion} concludes the paper.

\subsection{Notation}

Bold upper- and lower-case letters describe matrices and column vectors. Matrix transposition, conjugation, conjugate transposition and Frobenius norm are indicated respectively as $\mathbf{A}^{\mathrm{T}}$, $\mathbf{A}^{*}$, $\mathbf{A}^{\mathrm{H}}$ and $\|\mathbf{A}\|_F$. $[\mathbf{A}]_{ij}$ is the $(i,j)$-th entry of $\mathbf{A}$. With $\|\mathbf{a}\|_\mathbf{Q} = \mathbf{a}^\mathrm{H} \mathbf{Q} \mathbf{a}$ we denote the norm of $\mathrm{a}$ weighted by matrix $\mathbf{Q}$. $\mathrm{diag}(\mathbf{A})$ denotes the extraction of the diagonal of $\mathbf{A}$, while $\mathrm{diag}(\mathbf{a})$ is the diagonal matrix given by vector $\mathbf{a}$. $\mathrm{vec}(\mathbf{A})$ denotes the vectorization by columns of matrix $\mathbf{A}$. Symbols $\otimes$ and $\odot$ denote, respectively, the Kronecker and the Hadamard (element-wise) product between two matrices/vectors, while $\circledast \circledast$ is the 2D periodic convolution of signals.  $\mathbf{I}_n$ is the identity matrix of size $n$. Superscritps $\mathrm{TD}$, $\mathrm{FT}$ and $\mathrm{DD}$ denote the time-delay (slow-fast time), frequency-time and delay-Doppler domains, respectively. $|\mathcal{A}|$ denotes the cardinality of set $\mathcal{A}$. With  $\mathbf{a}\sim\mathcal{CN}(\boldsymbol{\mu},\mathbf{C})$ we denote a multi-variate circularly complex Gaussian random variable $\mathbf{a}$ with mean $\boldsymbol{\mu}$ and covariance $\mathbf{C}$. $\mathbb{E}[\cdot]$ is the expectation operator, while $\mathbb{R}$ and $\mathbb{C}$ stand for the set of real and complex numbers, respectively. $\delta_{n}$ is the Kronecker delta.

\begin{figure*}[!t] 
\centering
\includegraphics[width=\columnwidth]{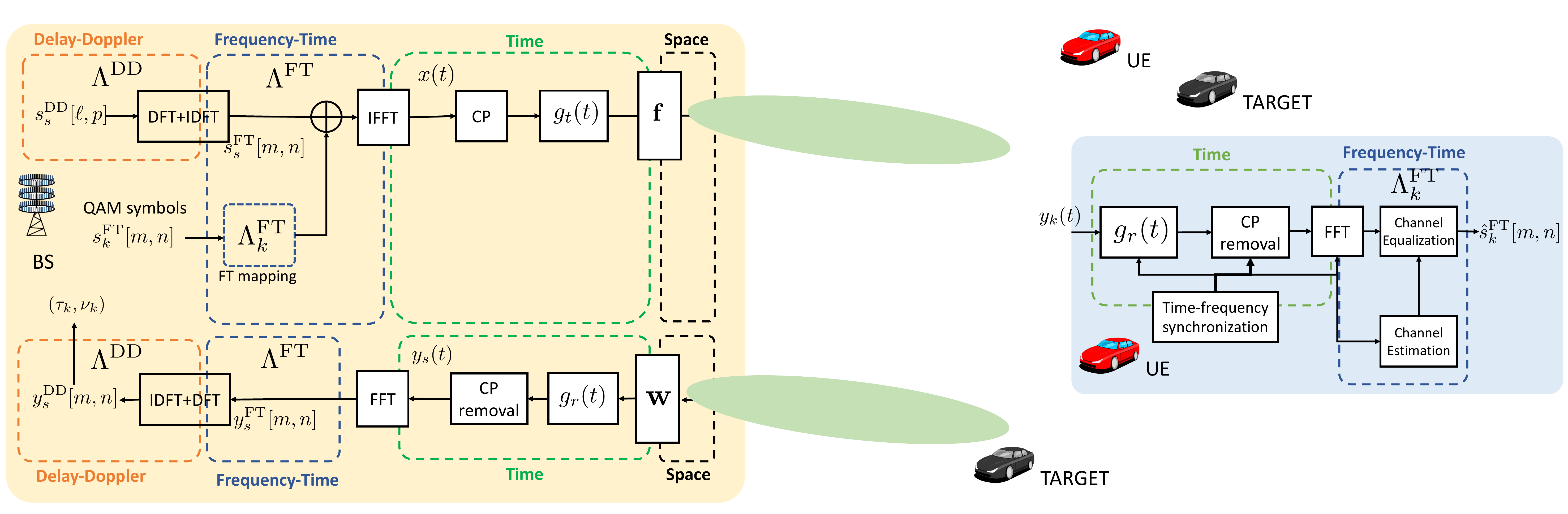}
\caption{Block scheme of the proposed dual-domain ISAC waveform: the BS employs the conventional OFDM scheme for communication, and superposes a sensing signal to detect the targets in the environment, estimating their delays and Doppler shifts in the DD domain. The Rx processing chain of the UE is not affected. }%: ISAC BS Tx (top-left) and sensing Rx (bottom-left); UE communication receiver (right)}
\label{fig:System Model}
\end{figure*}
%
%
%\begin{figure*}[b!]
%    \centering
%    \hrulefill
%   \begin{align}
 %   \mathbf{h}^\mathrm{FT}_{k}[m,n] &= T \,\Delta f \sum_{u=1}^U \alpha_{u,k} \,e^{j 2 \pi( \nu_{u,k} nT - m\Delta f \tau_{u,k})} G(m\Delta f)\,\mathbf{a}_L(\boldsymbol{\xi}_{u,k}), \label{eq:discrete_FT_comm_channel}
%    \\
%    \mathbf{H}^\mathrm{FT}_{s}[m,n] &= T \,\Delta f \sum_{q=1}^{Q} \beta_{q} \,e^{j 4 \pi( \nu_{q} nT - m\Delta f \tau_{q})} G(m\Delta f)\,\mathbf{a}_L(\boldsymbol{\vartheta}_{q})\mathbf{a}^\mathrm{H}_L(\boldsymbol{\vartheta}_{q}),\label{eq:discrete_FT_sensing_channel}
    %    \mathbf{h}^\mathrm{TD}_{k}(t,\tau) = \sum_{u=1}^{U} \alpha_{u,k} \, e^{j 2 \pi \nu_{u,k} t} \, \delta(\tau-\tau_{u,k}) \,\mathbf{a}_L(\boldsymbol{\xi}_{u,k})
%    \end{align}
%\end{figure*}
%
\section{Time-variant scenario}\label{sect:system_model}

%We consider an ISAC system where the ISAC node, i.e., BS, is equipped with one sub-array (panel) of $L$ antennas employed as a Tx and one panel of the same size as the sensing Rx (full-duplex operation). With a single antenna panel, the BS synthesizes a beam to serve $K$ single-antenna UEs while estimating their delay/range and Doppler/velocity through a properly designed ISAC waveform. We assume that \textit{(i)} the $K$ UEs are the only targets in the environment and \textit{(ii)} the $K$ UEs are scheduled on the scheduled on disjoint portions of the same FT resource grid. The first assumption avoids issues related to the target-to-UE association, easing the analytical developments for the waveform design. Notice that, in general, a BS has multiple panels; however, the single-panel assumption eases the analytical derivations herein.

Let us consider the ISAC system depicted in Fig.\ref{fig:System Model}, where the ISAC node, i.e., BS, is equipped with two antenna arrays of $L$ elements each, one employed as a Tx and one as the sensing Rx (full-duplex operation). The arrays can be either digital or analog. In the former case, the BS can synthesize optimal beampatterns in Tx and Rx for communication and sensing, as detailed in \cite{liu2018toward}. However, since the present analysis is independent of the specific implementation of the spatial precoder, we hereafter consider for simplicity two analog arrays. 
With the Tx array, the BS synthesizes an ISAC beam to serve $K$ single-antenna UEs while estimating the delay and Doppler of $Q$ targets. Without loss of generality, we assume that \textit{(i)} the $Q$ targets do not comprise the $K$ UEs \cite{liu2018beampattern} and \textit{(ii)} the $K$ UEs are scheduled on disjoint portions of the FT resource grid. 

In high-mobility scenarios, such as the vehicular one, the mmWave/sub-THz communication channel between the BS and the $k$-th UE is spatially sparse and doubly selective, characterized by a severe path loss frequencies and large Doppler shifts due to UEs' motion. 
%The discrete communication channel in the FT domain, modelled as the sum of $U$ paths \cite{samimi2016mmwave}, is expressed as \eqref{eq:discrete_FT_comm_channel},
The communication channel with the $k$-th UE in the time-delay (TD) domain is modelled as the sum of $U_k$ paths~\cite{samimi2016mmwave}
\begin{equation}\label{eq:channel_model_TD_comm}
   h^\mathrm{TD}_k(t,\tau) = \sum_{u=1}^{U_k} \alpha_{u,k} \, e^{j 2 \pi \nu_{u,k} t} \, g(\tau-\tau_{u,k})\, \zeta_{u,k}
\end{equation}
where: \textit{(i)} the amplitude of each path is $\alpha_{u,k}\sim\mathcal{CN}(0,\sigma^2_{u,k})$, with $\sigma^2_{u,k}\propto (f_0 R_{u,k})^{-2}$ (for distance $R_{u,k}$ and carrier frequency $f_0$); \textit{(ii)} $\nu_{u,k}$ is the Doppler shift for the $u$-th path; \textit{(iii)} $g(t)$ denotes the time response of the cascade of the pulse shaping and matched filters
\textit{(iv)} $\tau_{k}=R_{u,k}/c$ is the path delay \textit{(v)} $\zeta_{u,k}$ is the beamforming gain for the $k$-th UE, $u$-th path. Note that $\tau_{1,k}= R_k/v$ and $\nu_{1,k}= f_0 V_k/v$ are proportional to the range $R_k$ and radial velocity $V_k$ of the $k$-th target, where $v$ is the speed of light.

Similarly, the sensing channel is composed of the two-way path between the BS and all the $Q$ targets in the environment:
\begin{equation}\label{eq:channel_model_sensing_TD}
    h^\mathrm{TD}_\mathrm{sen}(t,\tau) = \sum_{q=1}^{Q} \beta_{q} \, e^{j 2 \pi \nu_{q} t} \, g(\tau-\tau_{q})\, \zeta_q
\end{equation}
% 
%\begin{figure*}
%\begin{equation}
    %\mathbf{H}^\mathrm{FT}_s(t,f) &= \sum_{q=1}^{Q} \beta_{q} \, e^{j 4 \pi \nu_{q} t} e^{-j 4 \pi \tau_{q} f} \, \mathbf{a}_L(\boldsymbol{\xi}_q) \mathbf{a}_L^{\mathrm{H}}(\boldsymbol{\xi}_q)\label{eq:channel_model_sensing_FT}\\
    %\mathbf{H}^\mathrm{DD}_s(\nu,\tau) = \sum_{q=1}^{Q} \beta_{q}\, \delta(\nu-2\nu_q)\, \delta(\tau-2\tau_{q}) \, \mathbf{a}_L(\boldsymbol{\xi}_q) \mathbf{a}_L^{\mathrm{H}}(\boldsymbol{\xi}_q)\label{eq:channel_model_sensing_DD}
%\end{equation}
%\hrulefill
%\end{figure*}
%
where $\beta_{q}\sim\mathcal{CN}(0,\sigma^2_{q})$, with $\sigma^2_{q}\propto
\Gamma_q f_0^{-2}R_{q}^{-4}$ is now the scattering amplitude of each target (comprising the radar cross section $\Gamma_q$), $\tau_{q}= 2R_q/v$, $\nu_{q}= 2 f_0 V_q/v$ and $\zeta_q$ is the beamforming gain for the sensing system. Notice that we are assuming that the channel includes only the two-way LOS path between the BS and each target. Multiple reflections are characterized by a higher path-loss (typically $\propto R^{-6}$) and are therefore not considered. 

The sensing channel \eqref{eq:channel_model_sensing_TD} can be further expressed in both frequency-time (FT) and delay-Doppler (DD) domains by Fourier transforms over delay and time dimensions, obtaining respectively \cite{Bello63}
%
%\begin{equation}
\begin{align}
    h^\mathrm{FT}_\mathrm{sen}(t,f) &= \sum_{q=1}^{Q} \beta_{q} \, e^{j 2 \pi \nu_{q} t} G(f) e^{-j 2 \pi \tau_{q} f} \, \zeta_q \label{eq:channel_model_sensing_FT}\\
    h^\mathrm{DD}_\mathrm{sen}(\nu,\tau) &= \sum_{q=1}^{Q} \beta_{q}\, \delta(\nu-\nu_q)\, g(\tau-\tau_{q}) \, \zeta_q \label{eq:channel_model_sensing_DD}
\end{align}
%\end{equation}
%
where $G(f)$ is the Fourier transform of $g(t)$. The same transformation can also be applied to the communication channel \eqref{eq:channel_model_TD_comm}.

\section{Transmitted ISAC Waveform}\label{sect:TxISAC}
%
%The transmitted ISAC signal is designed as the superimposition of two waveforms, the OFDM for communication and one for sensing, as depicted Fig.~\ref{fig:System Model}. The proper power scaling of the two allows \textit{(i)} to serve the intended UEs without suffering the distortion from the sensing signal at the UE side, and \textit{(ii)} to estimate delay and Doppler of each target without suffering the interference from the communication signal.
%
\begin{figure}[!t]
    \centering
    \includegraphics[width=0.65\columnwidth]{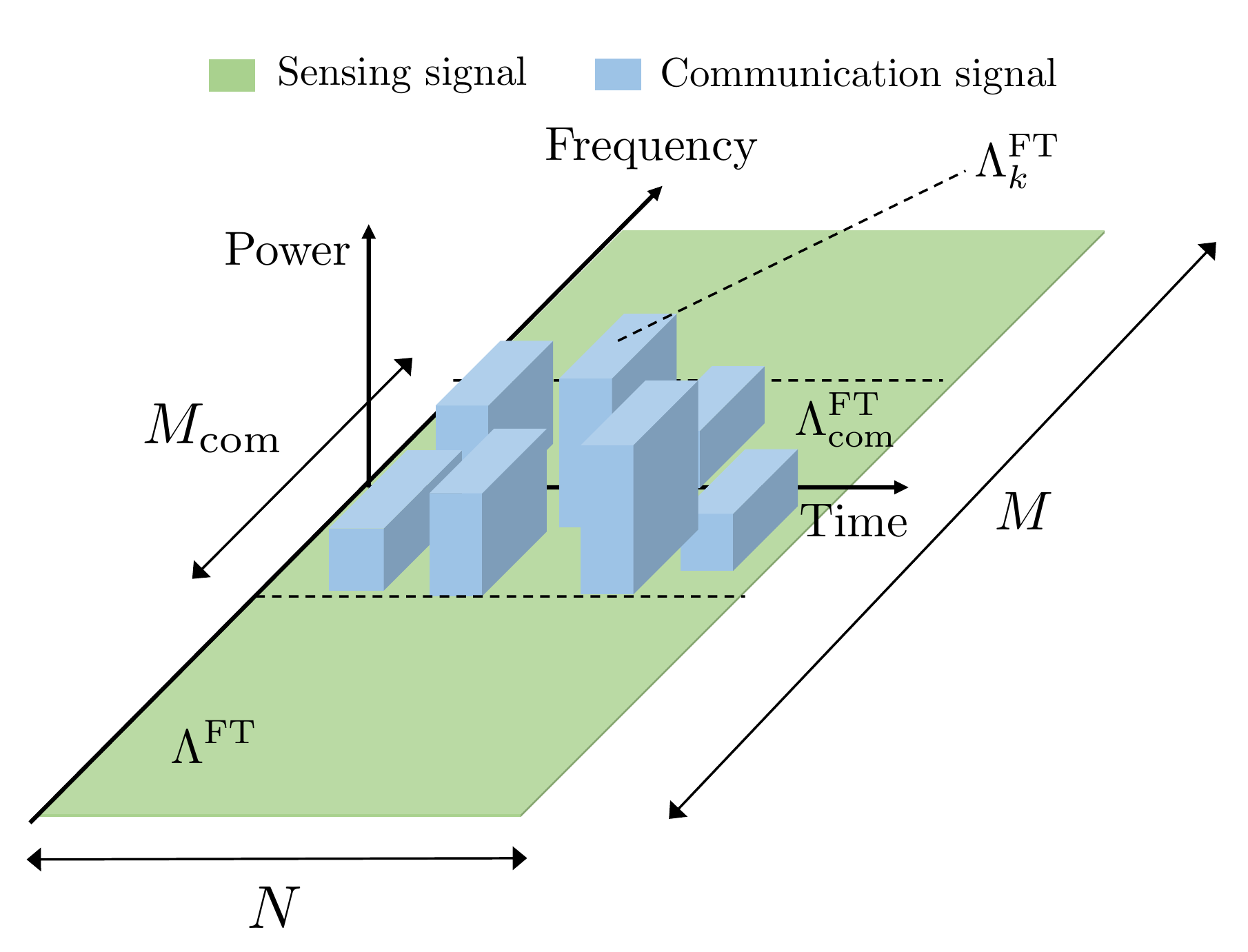}
    \caption{Resource allocation for communication and sensing signals in the proposed ISAC dual-domain system}
    \label{fig:resources}
\end{figure}

The transmitted ISAC waveform consists of the superposition of two signals, i.e., a communication signal, designed in the FT-domain and a sensing signal, designed in the DD-domain, as exemplified in Fig. \ref{fig:resources} with an example of resource allocation for communication and sensing. The proposed ISAC waveform exploits a certain pool of resources, $M$ in frequency/delay and $N$ in time/Doppler. In the FT domain, the set of resources can be represented by the grid
\begin{equation}\label{eq:TFgrid}
    \begin{split}
        \Lambda^{\mathrm{FT}} = \big\{ m\Delta f, nT \,\big|\, m =-\frac{M}{2},\dots,\frac{M}{2}-1,\,n =-\frac{N}{2},\dots,\frac{N}{2}-1 \big\},
    \end{split}
\end{equation}
where $\Delta f$ denotes the sub-carrier spacing and $T$ denotes the symbol duration of an OFDM symbol, comprising the cyclic prefix, i.e., $T=T'+T_{cp}$, with $\Delta f = 1/T'$. Reciprocally, the set of resources in the DD domain can be represented by grid 
\begin{equation}\label{eq:DDgrid}
    \begin{split}
        \Lambda^{\mathrm{DD}} = \big\{\ell\Delta \tau, p\Delta \nu\,\big|\, \ell =0 ,\dots,M-1, \,p = -\frac{N}{2},\dots,\frac{N}{2}-1\big\}
    \end{split}
\end{equation}
where $\Delta \tau = 1/(M\Delta f)$ and $\Delta \nu = 1/(NT)$ denote the delay and Doppler resolutions, respectively, which depend on the system bandwidth $(M\Delta f)$ and the downlink burst duration $(NT)$. The corresponding range and velocity resolutions are, respectively:
\begin{align}
    \Delta R = \frac{c}{2}\Delta \tau, \quad \quad \Delta V = \frac{c}{2 f_0 }\Delta \nu.
\end{align}
The Tx ISAC signal matrix in FT domain is designed as:
\begin{equation}
\begin{split}
    \mathbf{X}^{\mathrm{FT}} &= \boldsymbol{\Sigma}^\mathrm{FT}_\mathrm{com} \odot \mathbf{S}^{\mathrm{FT}}_\mathrm{com} + \sigma_\mathrm{sen}^\mathrm{FT} \mathbf{S}_\mathrm{sen}^{\mathrm{FT}},
\end{split}
\end{equation}
where $\mathbf{S}^{\mathrm{FT}}_\mathrm{com}\in\mathbb{C}^{M\times N}$ is the FT communication signal matrix, $\mathbf{S}^{\mathrm{FT}}_\mathrm{sen}\in\mathbb{C}^{M\times N}$ is the FT sensing signal matrix, $\boldsymbol{\Sigma}^\mathrm{FT}_\mathrm{com}\in\mathbb{R}_+^{M\times N}$ defines the square root of the allocated powers for the communication signal while $\sigma_\mathrm{sen}^\mathrm{FT}\geq 0$ is the amplitude of the sensing signal. 
The resulting bandwidth-integrated powers of communication and sensing signals are, respectively:
\begin{equation}
    P^\mathrm{FT}_\mathrm{com} = \frac{\|\boldsymbol{\Sigma}^\mathrm{FT}_\mathrm{com}\|^2_F}{N},
   \quad \quad
    P^\mathrm{FT}_\mathrm{sen} = M \left(\sigma_\mathrm{sen}^\mathrm{FT}\right)^2.
\end{equation}
As customary, the Tx signal $\mathbf{X}^{\mathrm{FT}}$ undergoes an $M$-point IFFT and a CP of duration $T_{cp}$ is appended before transmission. 
The notable characteristic of the proposed waveform is that the sensing processing, i.e., target detection, position, and velocity estimation, is carried out in the DD domain by means of an IDFT-DFT transformation $\mathbf{X}^{\mathrm{DD}} = \mathbf{F}_M^\mathrm{H}  \mathbf{X}^{\mathrm{FT}}\mathbf{F}_N$, where $\mathbf{F}_M\in\mathbb{C}^{M\times M}$ and $\mathbf{F}_N\in\mathbb{C}^{N\times N}$ are DFT matrices such that $\|\mathbf{F}_M\|_F=\sqrt{M}$ and $\|\mathbf{F}_N\|_F=\sqrt{N}$. When the sensing signal in FT domain is a 2D sinusoid occupying all the $MN$ resources (see Section \ref{subsect:sensing_signal} for details), the corresponding DD mapping is an impulse whose amplitude is augmented by $\sqrt{MN}$, i.e.,
\begin{equation}\label{eq:power_sensing_DD}
    \left(\sigma_\mathrm{sen}^\mathrm{DD}\right)^2 = \left(\sigma_\mathrm{sen}^\mathrm{FT}\right)^2 MN
\end{equation}
and the communication signal is regarded as a disturbance affecting the sensing processing with an average power on each DD bin that is
\begin{equation}\label{eq:power_comm_DD}
\left(\sigma_\mathrm{com}^{\mathrm{DD}}\right)^2 = \frac{\|\boldsymbol{\Sigma}^\mathrm{FT}_\mathrm{com}\|^2_F}{MN}.
\end{equation}
The proper tuning of the communication and sensing powers allows \textit{(i)} to serve the intended UEs without suffering the distortion from the sensing signal at the UE side and \textit{(ii)} to estimate delay and Doppler of each target without suffering the interference from the communication signal, increasing the resolution on delay estimation, as detailed in the following.

\subsection{Communication signal}\label{subsect:comm_signal}

%Let $M$ and $N$ be the total number of contiguous subcarriers and OFDM symbols in a downlink burst, respectively. The discrete FT resource grid employed for communication is therefore represented by the set  
%
%\begin{equation}\label{eq:TFgrid}
%    \begin{split}
%        \Lambda^{\mathrm{FT}} = \big\{ m\Delta f, nT \,\big|\, m &=-\frac{M}{2},\dots,\frac{M}{2}-1,\\n &=-\frac{N}{2},\dots,\frac{N}{2}-1 \big\} ,
%    \end{split}
%\end{equation}
%
%where $\Delta f$ and $T$ are the subcarrier spacing and the OFDM symbol duration, the latter comprising the cyclic prefix, i.e., $T=T'+T_{cp}$, with $\Delta f = 1/T'$. The resource grid can be also expressed as a Cartesian product of sets as $\Lambda^{\mathrm{FT}} =\mathcal{M} \times \mathcal{N}$, where $\mathcal{M}$ is the set of subcarriers and $\mathcal{N}$ is the set of OFDM symbols. 
For the $k$-th UE, the BS allocates a portion of the FT resources
\begin{equation}\label{eq:TFgrid_user}
    \begin{split}
        \Lambda_{k}^{\mathrm{FT}}= \big\{m\Delta f, nT \big\}\subseteq \Lambda^{\mathrm{FT}}
    \end{split}
\end{equation}
where $\Lambda_{k}^{\mathrm{FT}} \cap \Lambda_{\ell}^{\mathrm{FT}} =\emptyset$, for $k\neq \ell$ to avoid multi-user interference. For the purpose of this work, we also define the FT resource grid effectively occupied by the whole communication signal as:
\begin{equation}
    \Lambda^{\mathrm{FT}}_\mathrm{com} = \mathrm{cvxhull}\left(\bigcup_k \Lambda^{\mathrm{FT}}_k\right) 
\end{equation}
where $\mathrm{cvxhull}\left(.\right)$ is the discrete convex hull of the union of the $K$ allocated sets, as portrayed in Fig. \ref{fig:resources}. The available communication resources in $\Lambda^{\mathrm{FT}}_\mathrm{com}$ are $M_\mathrm{com} N$, where $M_\mathrm{com}$ is the maximum number of subcarriers allowed by bandwidth regulation (e.g., $3300$ subcarriers for $400$ MHz spectrum in the 5G NR frequency range (FR) 2 with $\Delta f = 120$ kHz). Usually, $M_\mathrm{com} < M$, in the order of 20-to-50\%, as detailed in Section \ref{sect:introduction}.
Within $\Lambda^{\mathrm{FT}}_\mathrm{com}$, the fraction of allocated resources for OFDM is defined as
\begin{equation}
    \eta = \dfrac{\big\lvert\bigcup_k \Lambda^{\mathrm{FT}}_k\big\rvert}{\big\lvert \Lambda^{\mathrm{FT}}_\mathrm{com}\big\rvert} = \dfrac{\big\lvert\bigcup_k \Lambda^{\mathrm{FT}}_k\big\rvert}{M_\mathrm{com} N},
\end{equation}
regulating the sparsity of the in-band allocated resources for communication.
The $k$-th UE communication signal in the FT domain is modelled as:
\begin{equation}\label{eq:Comm signal UE}
    \begin{split}
        \left[\mathbf{S}_k^{\mathrm{FT}}\right]_{m,n} = \begin{dcases}
          a_k[m,n] & \text{if} \,(m,n) \in \Lambda^\mathrm{FT}_{k}\\
          \quad 0 & \text{otherwise}
        \end{dcases},
   \end{split}
\end{equation}
where $\mathbf{S}_k^{\mathrm{FT}}\in\mathbb{C}^{M\times N}$ and $a_k[m,n]$ is a random information symbol drawn from a QAM constellation such that $\mathbb{E}[ a_k[m,n]a^*_\ell[m,n]]= \delta_{k-\ell}$. The overall communication signal can therefore be expressed as
\begin{equation}\label{eq:Comm signal}
    \begin{split}
        \mathbf{S}^{\mathrm{FT}}_\mathrm{com} = \sum_{k = 1}^{K} \mathbf{S}^{\mathrm{FT}}_k.
   \end{split}
\end{equation}
%
%where $\left(\sigma_{k}^{\mathrm{FT}}\right)^2$ denotes Tx power allocated for the $k$th UE.

%Therefore, the total power emitted by the BS for the communication with the $k$-th UE is 
%
%\begin{equation}
%    P_{k}^{\mathrm{FT}} = \frac{1}{N_k} \sum_{n\in\mathcal{N}_k}\sum_{m\in\mathcal{M}_k} \left(\sigma_{k}^{\mathrm{FT}}\right)^2 = M_k \left(\sigma_{k}^{\mathrm{FT}}\right)^2,
%\end{equation}
%
%summed over the subcarriers.

\subsection{Sensing signal}\label{subsect:sensing_signal}

The proposed sensing waveform is an impulse (or a combination of impulses) in the DD domain, located in $(\ell_i,p_i)$: 
\begin{equation}\label{eq:Sensing DD matrix multi imp}
    \left[\mathbf{S}_\mathrm{sen}^{\mathrm{DD}}\right]_{\ell,p} = \begin{dcases}
          1 & \text{if} \, \ell = \ell_i, p = p_i\\
          0 & \text{otherwise}
    \end{dcases},
\end{equation}
where $\mathbf{S}_\mathrm{sen}^{\mathrm{DD}}\in\mathbb{C}^{M\times N}$ is the DD sensing signal matrix.
The BS maps the sensing signal \eqref{eq:Sensing DD matrix multi imp} from the DD domain to the FT domain by means of a DFT-IDFT pair along the delay and Doppler dimensions, respectively, obtaining a 2D sinusoid~\cite{Bello63}:
\begin{equation}\label{eq:discreteISFFT}
    \mathbf{S}_\mathrm{sen}^{\mathrm{FT}} = \mathbf{F}_M  \mathbf{S}_\mathrm{sen}^{\mathrm{DD}}\mathbf{F}^\mathrm{H}_N
\end{equation}
The $(m,n)$-th entry of $\mathbf{S}_\mathrm{sen}^{\mathrm{FT}}$ is 
\begin{equation}\label{eq:discreteISFFT_sinusoid}
\begin{split}
[\mathbf{S}_\mathrm{sen}^{\mathrm{FT}}]_{m,n} = \frac{1}{\sqrt{M N}} e^{j 2 \pi \left(\frac{np_i}{N}-\frac{m\ell_i}{M}\right)}.
\end{split}
\end{equation}

\section{Received Signals}\label{sect:Rx_signal}

In the following, we detail the Rx communication signal at the UE side and the Rx sensing signal at the BS side, evaluating the signal-plus-distortion-and-noise ratio (SDNR) in both cases.

\subsection{Received communication signal at the UE} \label{subsect:received_signal_UE}

The signal received by the $k$-th UE in the FT domain, assuming perfect time-frequency synchronization, can be expressed as
\begin{equation}\label{eq:Rx_signal_UE}
\begin{split}
    \mathbf{Y}^\mathrm{FT}_k 
    = \mathbf{H}_k^\mathrm{FT} \odot \mathbf{X}^\mathrm{FT} + \mathbf{Z}^\mathrm{FT}_k =
    \underbrace{\mathbf{H}_k^\mathrm{FT} \odot \boldsymbol{\Sigma}^\mathrm{FT}_\mathrm{com} \odot \mathbf{S}^{\mathrm{FT}}_\mathrm{com}}_{\text{communication signal}\, \mathbf{B}^\mathrm{FT}_\mathrm{k,com}} 
    +
    \underbrace{\mathbf{H}_k^\mathrm{FT} \odot \sigma_\mathrm{sen}^\mathrm{FT} \mathbf{S}_\mathrm{sen}^{\mathrm{FT}}}_{\text{sensing signal} \,\mathbf{D}_\mathrm{k, sen}^\mathrm{FT}} + \mathbf{Z}^\mathrm{FT}_k
\end{split}
\end{equation}
where $\left[\mathbf{Z}^{\mathrm{FT}}_k\right]_{m,n} \sim\mathcal{CN}(0,\sigma_z^2)$ is an additive white noise affecting the $(m,n)$-th FT bin and
\begin{equation}\label{eq:discrete_FT_comm_channel}
\left[\mathbf{H}_k^\mathrm{FT}\right]_{m,n} = \sum_{u=1}^{U_k} \alpha_{u,k} \,e^{j 2 \pi( \nu_{u,k} nT - m\Delta f \tau_{u,k})} G(m\Delta f)\,\zeta_{u,k} \in \mathbb{C}^{M \times N}
\end{equation}
is $k$-th UE's FT discrete channel matrix in \eqref{eq:discrete_FT_comm_channel}. In \eqref{eq:Rx_signal_UE}, the communication signal accounts for the signal for all the $K$ UEs and the distortion due to the sensing signal is distinguished. The communication SDNR at the $(m,n)$-th FT bin of the $\Lambda_k^\mathrm{FT}$ grid is (before equalization):
\begin{equation}\label{eq:SDNR_comm_FT_gen}
\begin{split}
     \gamma^{\mathrm{FT}}_k 
     &= \frac{\mathbb{E}_{\Lambda_k^\mathrm{FT}}\left[\left\|\left[\mathbf{B}^\mathrm{FT}_\mathrm{k,com}\right]\right\|_F^2\right]}{\mathbb{E}_{\Lambda_k^\mathrm{FT}}\left[\left\|\left[\mathbf{D}^\mathrm{FT}_\mathrm{k,sen}\right]  \right\|_F^2\right] + \mathbb{E}_{\Lambda_k^\mathrm{FT}}\left[\left\|\left[\mathbf{Z}^\mathrm{FT}_\mathrm{k}\right]  \right\|_F^2\right]} 
\end{split}
\end{equation}
where $\mathbb{E}_{\Lambda_k^\mathrm{FT}}[\cdot]$ is the expectation operator taken over the support $\Lambda_k^\mathrm{FT}$, defined by the FT resources allocated for the $k$-th UE. The upper bound of \eqref{eq:SDNR_comm_FT_gen} can be computed in closed form as
\begin{equation}\label{eq:SDNR_comm_FT}
\begin{split}
     \gamma^{\mathrm{FT}}_k \leq \frac{\kappa^2 \left(\sigma_k^\mathrm{FT}\right)^2}{\kappa^2 \left(\sigma_\mathrm{sen}^\mathrm{FT}\right)^2 + \sigma_z^2}
\end{split}
\end{equation}
where $\kappa^2 = \sum_{u=1}^{U_k} \sigma_{u,k}^2 L$ denotes the maximum channel gain for the $k$-th UE and $\sigma_k^\mathrm{FT}$ is the entry of $\boldsymbol{\Sigma}^\mathrm{FT}_\mathrm{com}$ corresponding to the $k$-th UE, assuming equal power allocation across resources. After time-frequency synchronization, the sensing signal in FT domain is known at the UE side, which can possibly operate the cancellation of $\mathbf{D}_\mathrm{k,sen}^{FT}$ obtaining a higher SNR bound, where only $\sigma_z^2$ is left at the denominator of \eqref{eq:SDNR_comm_FT}. However, the ISAC system can be designed to avoid the cancellation procedure at the UE reducing the Rx complexity. The achievable rate of the $k$-th DL communication link is  
\begin{equation}\label{eq:capacity_dualdomain}
    C =\frac{T'}{T'+T_{cp}} \, \log_2 \left(1 + \gamma^{\mathrm{FT}}_{k}\right)\,\,\text{[bps/Hz]}
\end{equation}
where the first term accounts for the loss due to the presence of CP.
\subsection{Received sensing signal at the BS}\label{subsect:received_signal_BS}

The Rx signal at the BS from the two-way propagation in the environment is directly dependent on the chosen BF direction at the Rx panel. In the most general case, the Rx signal in the FT domain can be expressed as
\begin{equation}\label{eq:Rx_signal_BS}
\begin{split}
    \mathbf{Y}_\mathrm{sen}^\mathrm{FT} 
    = \mathbf{H}_\mathrm{sen}^\mathrm{FT} \odot \mathbf{X}^\mathrm{FT} + \mathbf{Z}^\mathrm{FT}_\mathrm{sen}= \underbrace{\mathbf{H}_\mathrm{sen}^\mathrm{FT} \odot \sigma_\mathrm{sen}^\mathrm{FT} \mathbf{S}_\mathrm{sen}^{\mathrm{FT}}}_{\text{sensing signal}\, \mathbf{D}_\mathrm{sen}^\mathrm{FT}}
    + \underbrace{\mathbf{H}_\mathrm{sen}^\mathrm{FT} \odot \boldsymbol{\Sigma}^\mathrm{FT}_\mathrm{com} \odot \mathbf{S}^{\mathrm{FT}}_\mathrm{com}}_{\text{communication signal}\, \mathbf{B}^\mathrm{FT}_\mathrm{com}}
    + 
    \mathbf{Z}^\mathrm{FT}_\mathrm{sen}
\end{split}
\end{equation}
where 
\begin{equation}
\left[\mathbf{H}_\mathrm{sen}^\mathrm{FT}\right]_{m,n} = \sum_{q=1}^{Q} \beta_{q} \,e^{j 2 \pi( \nu_{q} nT - m\Delta f \tau_{q})} G(m\Delta f)\,\zeta_q,\label{eq:discrete_FT_sensing_channel}\in \mathbb{C}^{M\times N}
\end{equation}
is the discrete FT sensing channel matrix in \eqref{eq:discrete_FT_sensing_channel}, and the noise term $\left[\mathbf{Z}_\mathrm{sen}^\mathrm{FT}\right]_{m,n}\sim\mathcal{CN}(0,L \sigma^2_z)$ in \eqref{eq:Rx_signal_BS} is obtained after Rx BF.

\textbf{Remark:} While the communication channel $\mathbf{H}_k^\mathrm{FT}$ is considered \textit{after} the usual time-frequency synchronization at the Rx, the sensing channel $\mathbf{H}_\mathrm{sen}^\mathrm{FT}$ does not include the effect of the synchronization, retaining the true delay and Doppler shifts of the targets. This modelling assumption is coherent with a practical system where Tx and Rx are perfectly clock-synchronized (as for radars). It is important to remark that the signal in \eqref{eq:Rx_signal_BS} is obtained by direct sampling the continuous-time signal after the matched filtering, discarding the CP and operating an $M$-point DFT. This processing chain is valid if the CP includes all the sensing echoes, thus $T_{cp} \geq 2 R_{max}/v$, otherwise other methods shall be used (e.g., cross-correlation with the TD Rx signal).

The sensing processing is operated in the DD domain, hence, the received signal in DD can be expressed as 
\begin{equation}\label{eq:Rx_signal_BS_DD}
\begin{split}
    \mathbf{Y}_\mathrm{sen}^\mathrm{DD} 
    = \mathbf{F}_M \, \mathbf{Y}_\mathrm{sen}^\mathrm{FT} \, \mathbf{F}_N^\mathrm{H}  
    = \underbrace{\mathbf{H}_\mathrm{sen}^\mathrm{DD} \circledast \circledast \, \sigma_\mathrm{sen}^\mathrm{DD} \mathbf{S}_\mathrm{sen}^{\mathrm{DD}}}_{\text{sensing signal}\, \mathbf{D}_\mathrm{sen}^\mathrm{DD}(\boldsymbol{\tau}, \boldsymbol{\nu})}
    +
    \underbrace{\mathbf{H}_\mathrm{sen}^\mathrm{DD} \circledast \circledast \, \left(\boldsymbol{\Sigma}^\mathrm{DD}_\mathrm{com} \odot \mathbf{S}^{\mathrm{DD}}_\mathrm{com}\right)  }_{\text{communication signal}\, \mathbf{B}^\mathrm{DD}_\mathrm{com}}
    + 
    \mathbf{Z}^\mathrm{DD}_\mathrm{sen}
\end{split}
\end{equation}
where $\mathbf{H}_\mathrm{sen}^\mathrm{DD}$ is the discrete DD sensing channel matrix whose entries are:
\begin{equation}\label{eq:discrete_DD_sensing_channel}
\begin{split}
\left[\mathbf{H}_\mathrm{sen}^\mathrm{DD}\right]_{\ell,p} =  \frac{1}{\sqrt{MN}}\sum_{q=1}^{Q} \beta_{q} e^{- j 2 \pi \nu_q \tau_q}\, \frac{\sin\left(\pi(p - \nu_{q}/\Delta \nu)\right)}{\sin\left(\frac{\pi}{N}(p - \nu_{q}/\Delta \nu)\right)}\, g(\ell -\tau_{q}/\Delta\tau)\, \zeta_q.
\end{split}
\end{equation}
In \eqref{eq:Rx_signal_BS_DD}, the sensing signal $\mathbf{D}_\mathrm{sen}^\mathrm{DD}(\boldsymbol{\tau}, \boldsymbol{\nu})$ includes all the $Q$ targets delays $\boldsymbol{\tau}$ and Dopplers $\boldsymbol{\nu}$. The general expression of the maximum likelihood estimator (MLE) is
\begin{equation}\label{eq:MLE_gen}
(\widehat{\boldsymbol{\tau}},\widehat{\boldsymbol{\nu}}) = \underset{\boldsymbol{\tau},\boldsymbol{\nu} }{\mathrm{argmin}} \left(\left\|\mathbf{y}_\mathrm{sen}^\mathrm{DD} - \mathbf{d}_\mathrm{sen}^\mathrm{DD}\right\|^2_{\mathbf{C}_n}\right),
\end{equation}
where $\mathbf{y}_\mathrm{sen}^\mathrm{DD} = \mathrm{vec}\left( \mathbf{Y}_\mathrm{sen}^\mathrm{DD}\right)$, $\mathbf{d}_\mathrm{sen}^\mathrm{DD} = \mathrm{vec}\left( \mathbf{D}_\mathrm{sen}^\mathrm{DD}\right)$, and $\mathbf{C}_n$ is the covariance matrix of the noise plus communication signal $\mathbf{n}_\mathrm{sen}^\mathrm{DD} =\mathrm{vec}\left(\mathbf{B}^\mathrm{DD}_\mathrm{com} + \mathbf{Z}^\mathrm{DD}_\mathrm{sen}\right)$. For a single target $Q = 1$ and a diagonal covariance matrix $\mathbf{C}_n = \sigma_n^2 \mathbf{I}_{MN}$, the MLE over the DD grid simplifies to the 2D periodogram \cite{BigS}.
%
%\begin{equation}\label{eq:MLE_delay_doppler}
%    (\widehat{\tau},\widehat{\nu}) = \underset{\tau,\nu }{\mathrm{argmax}} \left(\left|\mathbf{Y}_\mathrm{sen}^\mathrm{DD}\right|^2\right),
%\end{equation}
%
However, in general, the multiple peaks selection, herein adopted, is sub-optimal due to the coupling between the $Q$ targets responses and the colored noise, i.e., non-diagonal $\mathbf{C}_n$. Remarkably, using a single sensing impulse in the DD domain allows the exploration of the echoes of the $Q$ targets without additional processing. The sensing performance for detecting the $q$-th target depends on the sensing SDNR $\gamma_\mathrm{q,sen}^\mathrm{DD}$, computed as follows:
    \begin{equation}\label{eq:SNR_DD_q}
        \begin{split}
        \gamma_\mathrm{q,sen}^\mathrm{DD} = \frac{\mathbb{E}\left[\left|\left[\mathbf{D}_\mathrm{q,sen}^\mathrm{DD} \right]_{\tilde{\ell}_q,  \tilde{p}_q}\right|^2\right]}{\sum_{\substack{j = 1 \\ j \neq q}}^Q \mathbb{E}\left[\left|\left[\mathbf{D}_\mathrm{j,sen}^\mathrm{DD} \right]_{ \tilde{\ell}_q,  \tilde{p}_q}\right|^2\right] + \mathbb{E}\left[\left|\left[\mathbf{B}_\mathrm{com}^\mathrm{DD} \right]_{ \tilde{\ell}_q, \tilde{p}_q}\right|^2\right] + \mathbb{E}\left[\left|\left[\mathbf{Z}_\mathrm{sen}^\mathrm{DD} \right]_{ \tilde{\ell}_q,  \tilde{p}_q}\right|^2\right]}
        \end{split}
    \end{equation}
where $\tilde{\ell}_q = (\tau_q / \Delta\tau) + \ell_i$, and $\tilde{p}_q = (\nu_q /\Delta\nu) + p_i$ are the delay and Doppler indexes of the received $q$-th target pulse, respectively, while the sensing signal for the $q$-th target $\mathbf{D}_\mathrm{q,sen}^\mathrm{DD}(\tau_q, \nu_q)$ can be expressed as 
    \begin{equation}\label{eq:qth_sensing_sig_DD}
        \mathbf{D}_\mathrm{q,sen}^\mathrm{DD}(\tau_q, \nu_q) = \frac{\sigma_\mathrm{sen}^\mathrm{DD}}{\sqrt{MN}} \beta_q e^{- j 2 \pi \nu_q \tau_q}\, \frac{\sin\left(\pi(p - \nu_{q}/\Delta \nu - p_i)\right)}{\sin\left(\frac{\pi}{N}(p - \nu_{q}/\Delta \nu - p_i)\right)} g(\ell - \tau_{q}/\Delta\tau - \ell_i) \, \zeta_q.
    \end{equation}
The upper bound of the SDNR $\gamma_\mathrm{q,sen}^\mathrm{DD}$ in \eqref{eq:SNR_DD_q} can be computed as
\begin{equation}\label{eq:SNR_DD_SEN_UB}
    \gamma_\mathrm{q,sen}^\mathrm{DD} \leq \frac{\kappa_\mathrm{q,sen}^2 \, \left(\sigma_\mathrm{sen}^\mathrm{DD}\right)^2}{\sum_{\substack{j = 1 \\ j \neq q}}^Q \kappa_\mathrm{j,sen}^2 \, \chi_{q,j} \, \left(\sigma_\mathrm{sen}^\mathrm{DD}\right)^2 + \kappa_\mathrm{sen}^2 \left(\sigma_\mathrm{com}^\mathrm{DD}\right)^2 + \sigma_z^2}
\end{equation}
where $\kappa_\mathrm{q,sen}^2 = \sigma_q^2 \, L^2$, and $\kappa_\mathrm{sen}^2 = \sum_{q=1}^Q \kappa_\mathrm{q,sen}^2$. The term $\chi_{q,j}$ accounts for the coupling between the $q$-th and $j$-th targets and it is defined as
\begin{equation}
    \chi_{q,j} = \frac{1}{(MN)^2}
    \left|\frac{\sin\left(\pi((\nu_{q}-\nu_{j})/\Delta \nu) \right)} {\sin\left(\frac{\pi}{N}((\nu_{q}-\nu_{j})/\Delta \nu)\right)}  g\left(\frac{\tau_{q} - \tau_j}{\Delta\tau}\right)\right|^2 \leq 1,
\end{equation}
where the equality holds only when the $j$-th target has the same delay and Doppler of the $q$-th target.
Now, the key observation is that the sensing SDNR in DD $\gamma_\mathrm{q,sen}^\mathrm{DD}$ embeds the \textit{processing gain} $MN$ (as expressed by \eqref{eq:power_sensing_DD} in Section \ref{sect:TxISAC}), thus it is directly proportionally to the employed bandwidth $M\Delta f$ and observation time $NT$.
The processing gain allows mitigating the increased path-loss of sensing ($\propto R^{-4}$ compared to $\propto R^{-2}$ of the communication links) as well as the disturbance from the OFDM communication signal. The power allocation, to enable proper communication and sensing operations, is formalized as an optimization problem in the following section.

\section{ISAC Power Allocation}\label{sect:power_opt}
\begin{figure}
    \centering
    \includegraphics[width=0.5\columnwidth]{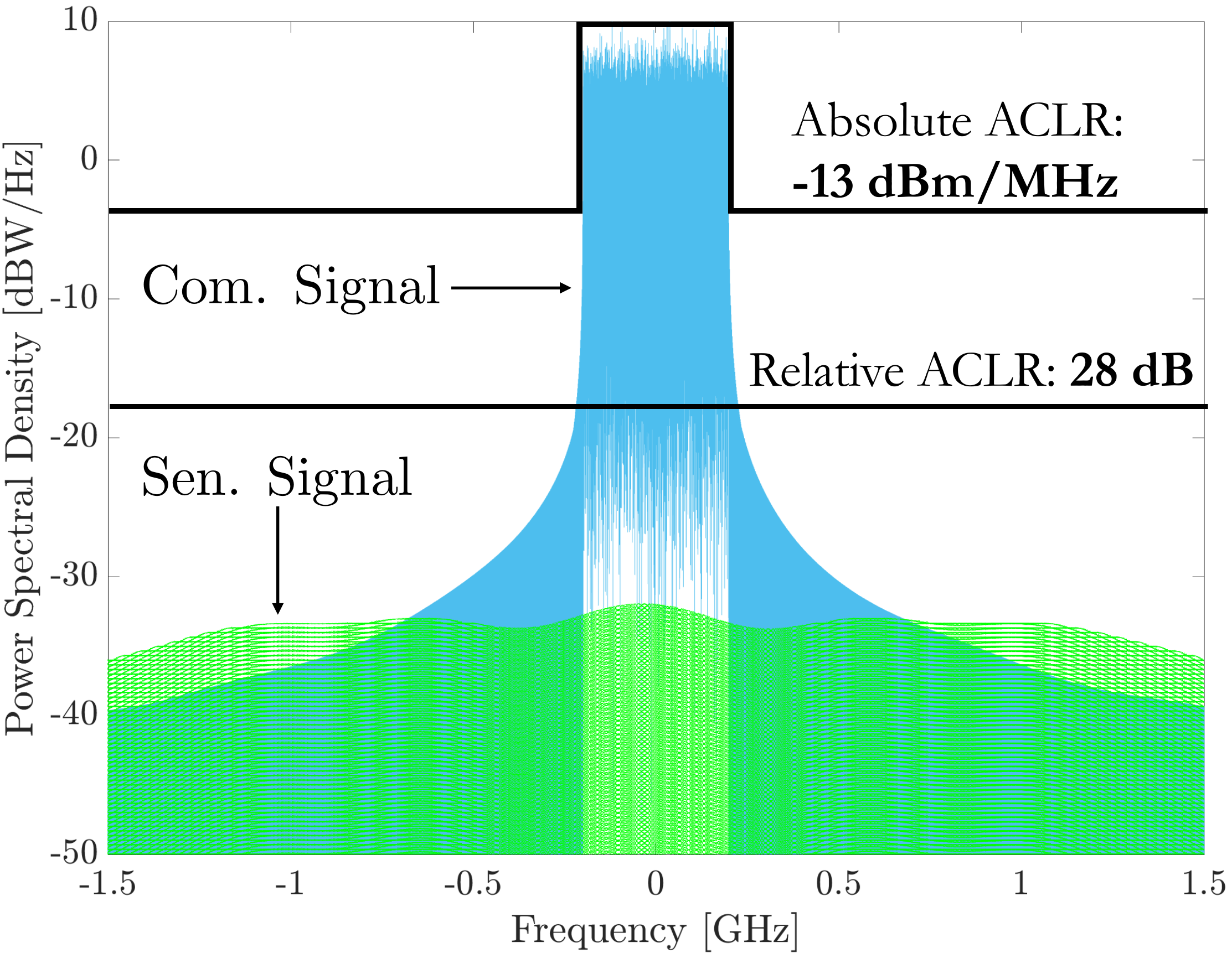}
    \caption{Example of superposition of a sensing signal to the legacy OFDM one. The former shall fall below the absolute (on the power spectral density) and relative (on the bandwidth integrated power) ACLR limits. }
    \label{fig:ACLR_limit}
\end{figure}

The proposed ISAC waveform allows the sensing signal to exceed the regulatory spectrum for communications to increase the delay/range resolution. In other words, when the resource grid used for sensing is significantly larger than the one set for communication, i.e., $\Lambda^\mathrm{FT}_\mathrm{com}\subset\Lambda^\mathrm{FT}$, $M_\mathrm{com}< M$, the range resolution can be enhanced provided that the sensing signal does not exceed a given bandwidth-integrated power threshold expressed in terms of ACLR limit~\cite{ETSI_BSrequirements}. The problem is graphically portrayed in Fig. \ref{fig:ACLR_limit}. The proposed ISAC waveform design consists of optimizing the power allocation while \textit{(i)} minimizing the total emitted power \textit{(ii)} ensuring a desired communication and sensing performance and \textit{(iii)} enforcing regulatory constraints on the out-of-band (OB) radiation. For instance, BS employed for 5G NR FR2 systems (\textit{BS type 2-O}) are manufactured to work within the $24.25-33.4$ GHz spectrum, on a typical $3$ GHz of bandwidth, while the maximum bandwidth for communication is fixed to $400$ MHz for single carrier and up to $1600$ MHz with carrier aggregation \cite{TS_38101}.

The power allocation problem can be formulated as:
\begin{subequations}
\begin{alignat}{2}
& \underset{\sigma_\mathrm{sen}^\mathrm{FT},\boldsymbol{\Sigma}_\mathrm{com}^\mathrm{FT}}{\mathrm{minimize}}     & \qquad & P_{\mathrm{tot}}^{\mathrm{FT}} = P^\mathrm{FT}_{\mathrm{ib}} + P^\mathrm{FT}_{\mathrm{ob}} \label{eq:prob3}\\
&\mathrm{subject\, to} &      &\gamma^\mathrm{FT}_k \geq \gamma^\mathrm{FT}_\mathrm{thr} \qquad\quad\,\,\, k=1,...,K \label{eq:constraint1_prob3}\\
&                  &      &  \gamma^\mathrm{DD}_\mathrm{q,sen} \geq \gamma^\mathrm{DD}_\mathrm{thr},\qquad\quad q=1,...,Q\label{eq:constraint2_prob3}\\
&                  &      &  \frac{P_{\mathrm{ib}}^{\mathrm{FT}}}{P^\mathrm{FT}_{\mathrm{ob}}} \geq \mathrm{ACLR}_\mathrm{rel}\label{eq:constraint3_prob3} \\ 
&                  &      & \left(\sigma_\mathrm{sen}^\mathrm{FT}\right)^2 \leq \mathrm{ACLR}_\mathrm{abs} \Delta f \label{eq:constraint4_prob3}\\
&                  &      & [\boldsymbol{\Sigma}_\mathrm{com}^{\mathrm{FT}}]_{ij} \geq 0 \qquad\quad \forall i,j\label{eq:constraint5_prob3}\\
&                  &      & \sigma_\mathrm{sen}^{\mathrm{FT}} \geq 0 \label{eq:constraint6_prob3}\\
&                  &      & P_{\mathrm{tot}}^{\mathrm{FT}} \leq P_{\mathrm{max}}^{\mathrm{FT}} \label{eq:constraint7_prob3}
\end{alignat}
\end{subequations}
where $P^\mathrm{FT}_{\mathrm{ob}} = M_\mathrm{ob}\left(\sigma_\mathrm{sen}^\mathrm{FT}\right)^2$ is the OB emitted power with $M_\mathrm{ob}$ denoting the number of OB sub-carriers, and 
\begin{equation}
    P^\mathrm{FT}_\mathrm{ib} = P^\mathrm{FT}_\mathrm{com} + M_\mathrm{com} \left(\sigma_\mathrm{sen}^\mathrm{FT}\right)^2
\end{equation} 
denotes the in-band (IB) emitted power, which consists of communication and sensing powers.
In \eqref{eq:prob3}, constraints \eqref{eq:constraint1_prob3} and \eqref{eq:constraint2_prob3} set the desired communication and sensing performance in terms of SDNR in FT and DD domains, respectively. For instance, $\gamma^\mathrm{FT}_\mathrm{thr}$ can be set according to a required data rate by the UEs, while $\gamma^\mathrm{DD}_\mathrm{thr}$ shall guarantee a pre-defined probability of correct detection \cite{richards2005fundamentals}. Constraints \eqref{eq:constraint3_prob3} and \eqref{eq:constraint4_prob3} limit the relative OB radiation power w.r.t. IB and the absolute power, measured with $\mathrm{ACLR}_\mathrm{rel}$ [dB] power gap and $\mathrm{ACLR}_\mathrm{abs}$ (usually specified in terms of [dBm/MHz]). Constraint  \eqref{eq:constraint7_prob3} sets the total power budget. The problem in \eqref{eq:prob3} is convex and can be cast within the geometric programming framework, and easily solvable \cite{boyd2007GPtut}.

\begin{figure}[!b]
    \centering
     \includegraphics[width=0.6\columnwidth]{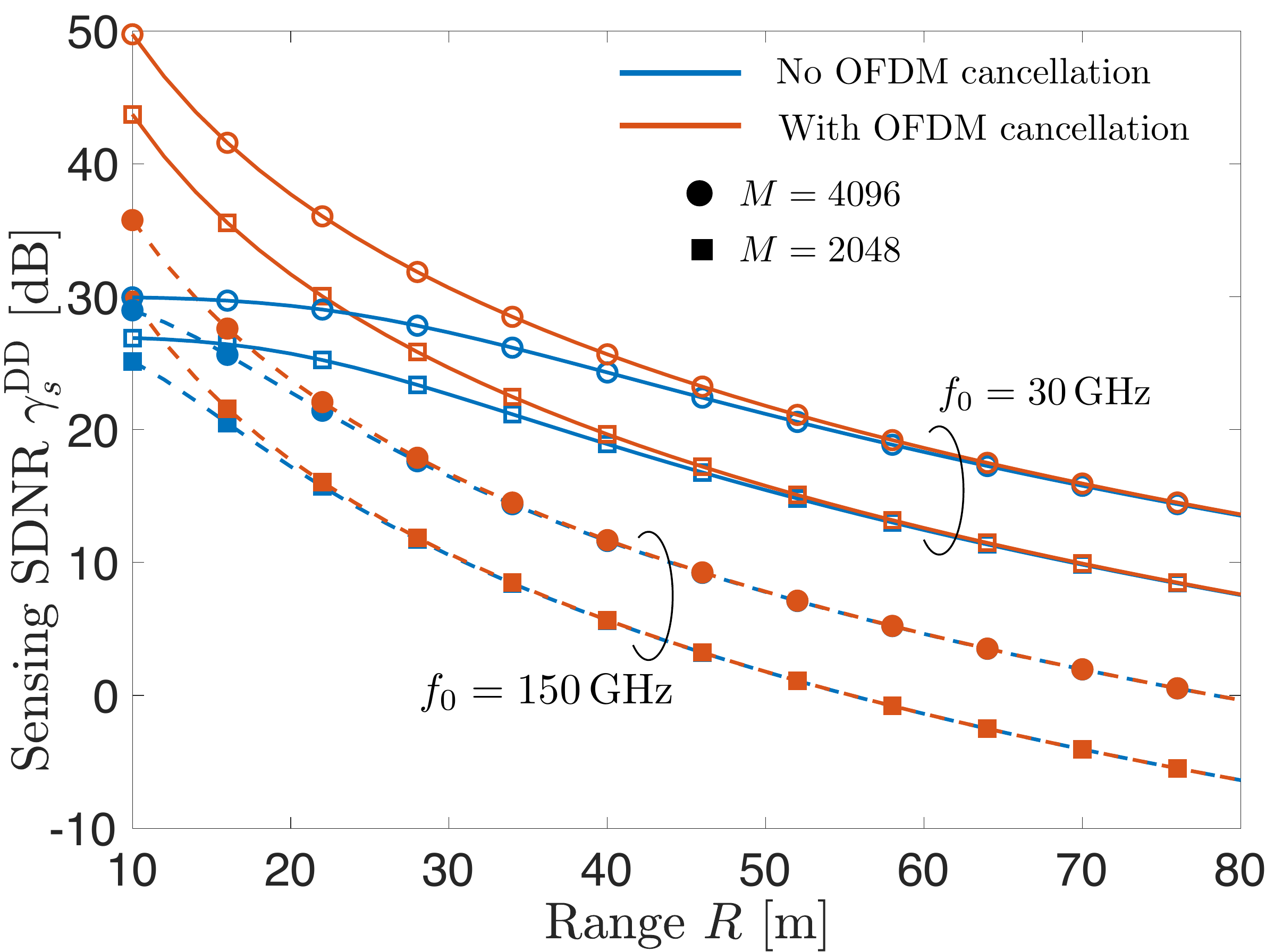}
    \caption{Sensing SNR in the DD domain at the BS, after the IDFT+DFT (matched filter). For a fixed bandwidth $B=1$ GHz, the effective coverage range for the dual domain ISAC waveform depends on the number of subcarriers $M$. The OFDM cancellation allows improving the performance only in very near range and for comparatively low carrier frequencies, while it is almost useless in other settings. }
    \label{fig:SNR_DD_range}
\end{figure}

\section{Performance Metrics and Numerical Results}\label{sect:numerical_results}

\begin{table}[t!]
    \centering
    \caption{Simulation Parameters}
    \begin{tabular}{l|c|c}
    \toprule
        \textbf{Parameter} &  \textbf{Symbol} & \textbf{Value(s)}\\
        \hline
        Carrier frequency & $f$  & $30$ GHz \\
        Bandwidth & $B$ & $1$ GHz\\
        Transmitted power & $P^\mathrm{FT}_{tot}$ & $43$ dBm\\
        Number of antennas &$L$& $100$\\
        Number of subcarriers & $M$ & $1024$\\
        Number of symbols & $N$ & $128$\\
        Symbol duration & $T'$ & $1.024$ $\mu$s \\
        CP & $T_{cp}$ & $0.102$ $\mu$s\\
        Sensing-to-comm FT ratio & $(\sigma^\mathrm{FT}_\mathrm{sen}/\sigma^\mathrm{FT}_\mathrm{com})^2$ & $10^{-3}$\\
        OFDM occupancy ratio & $\eta$ & $50$ \%\\
         Range & $R$ & $50$ m\\
        \bottomrule
    \end{tabular}
    \label{tab:SimParam}
\end{table}

This section validates the performance of the proposed ISAC design by presenting numerical results. We compare the dual domain waveform with the conventional OFDM and the OTFS in terms of: \textit{(i)} CRB on delay and Doppler estimation of targets; \textit{(ii)} MTER of the ambiguity function \textit{(iii)} achievable rate at the UEs. The CRB of the OTFS is derived in \cite{gaudio2020_OFDM_vs_OTFS}, and is not reported here for brevity, while the CRB of the dual-domain and the OFDM is reported in Appendix \ref{app:CRB} for one and two targets. The MTER, derived from the ambiguity function of the Tx ISAC waveform (Appendix \ref{app:ambiguity}), measures the fractional energy ratio of the main lobe w.r.t. the total, being therefore an indicator of delay-Doppler resolution of the ISAC system. The achievable rate of dual-domain and OFDM waveforms is evaluated by using \eqref{eq:capacity_dualdomain} (where we use the SNR upper-bound for OFDM), while its expression for OTFS is again derived in \cite{gaudio2020_OFDM_vs_OTFS}. Unless otherwise noted, we use the parameters in Table \ref{tab:SimParam} for simulation parameters. 

\subsection{ISAC performance with single target }

The first result we report is aimed at determining the effective sensing range of the dual-domain ISAC signal. Fig. \ref{fig:SNR_DD_range} shows the sensing SDNR in DD domain at the BS $\gamma_\mathrm{sen}^\mathrm{DD}$ varying the BS-target range $R$. We consider one target/UE, two carrier frequencies $f_0=30,150$ GHz and $M=2048,4096$ subcarriers. Here, the OFDM communication signal occupies half of the available bandwidth ($M_\mathrm{com}/M = 0.5$) and with 50\% resource occupancy ($\eta=0.5$). Blue curves show the SDNR, while red curves the SNR (with perfect OFDM signal cancellation at the BS Rx side). As expected, at lower frequencies, the Rx sensing signal in the near range is severely affected by the communication signal, thus a proper cancellation at BS would enhance the performance. It is interesting to notice that, for fixed bandwidth, the processing gain increases with $M$, thus it is practically ruled by the subcarriers' spacing $\Delta f$. This allows trading between the complexity of the Rx processing at the BS and the performance, still employing a DFT-based approach. 

\begin{figure}[!t]
    \centering
    \subfloat[][Root CRB comparison (single target)]{\includegraphics[width=0.6\columnwidth]{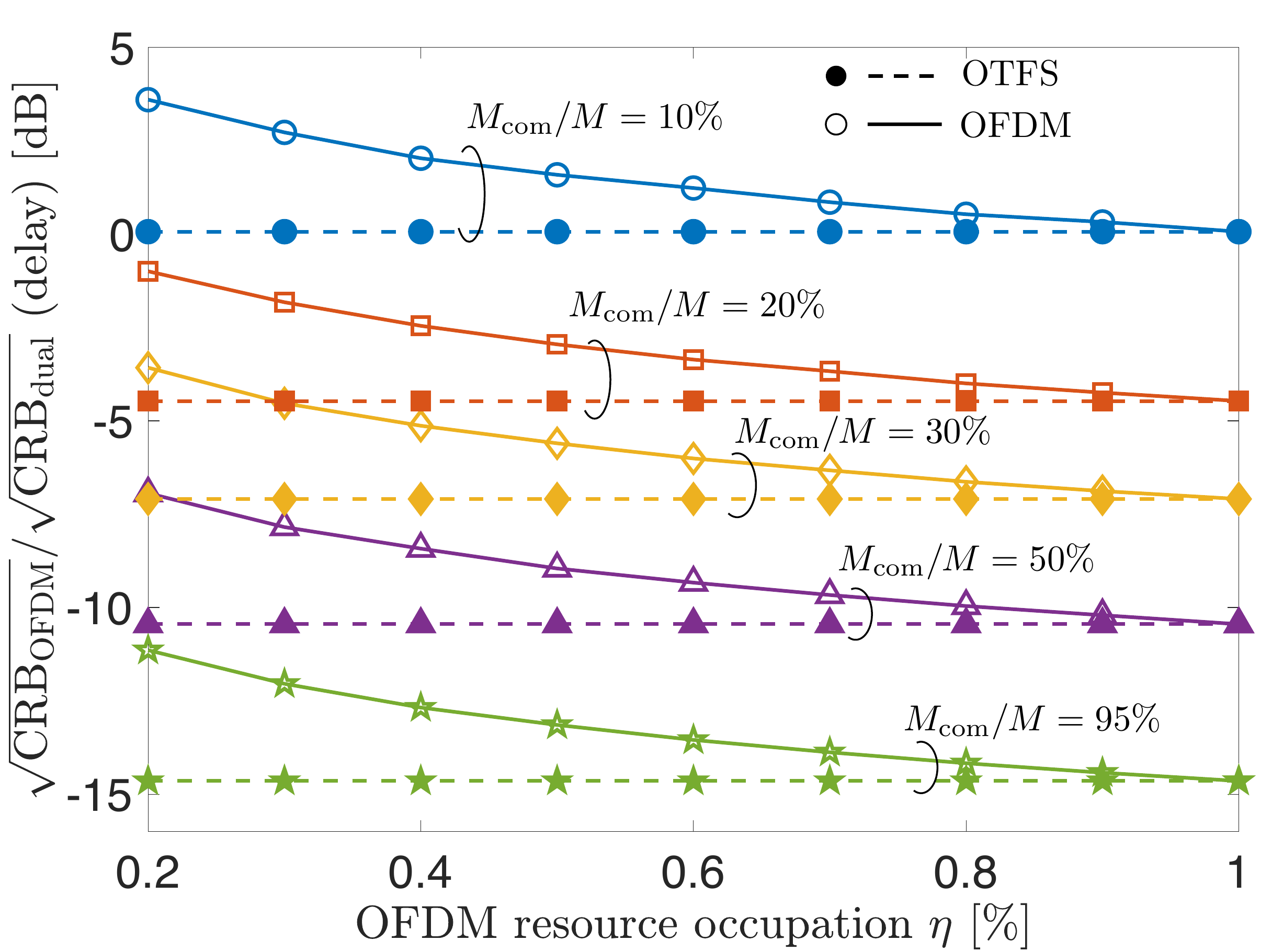}\label{subfig:CRB_delay}}\\
    \subfloat[][MTER]{\includegraphics[width=0.6\columnwidth]{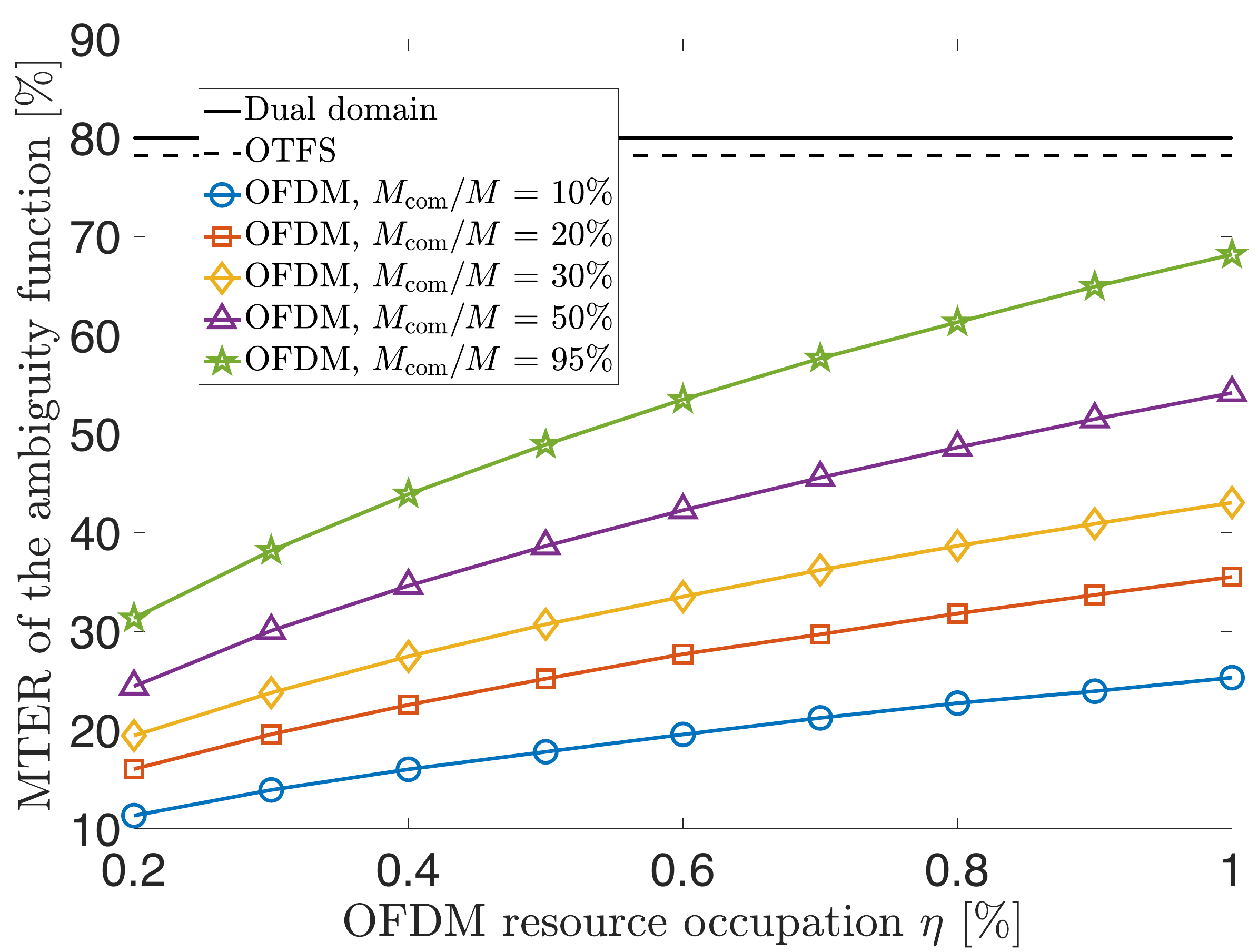}\label{subfig:MSLR}}
    \caption{Sensing performance comparison between dual-domain, OFDM and OTFS waveforms, in terms of (\ref{subfig:CRB_delay}) root CRB ratio on delay estimation (single target \ref{subfig:MSLR}) MTER. }
    \label{fig:Performance}
\end{figure}

\begin{figure}[!t]
    \centering
    \includegraphics[width=0.6\columnwidth]{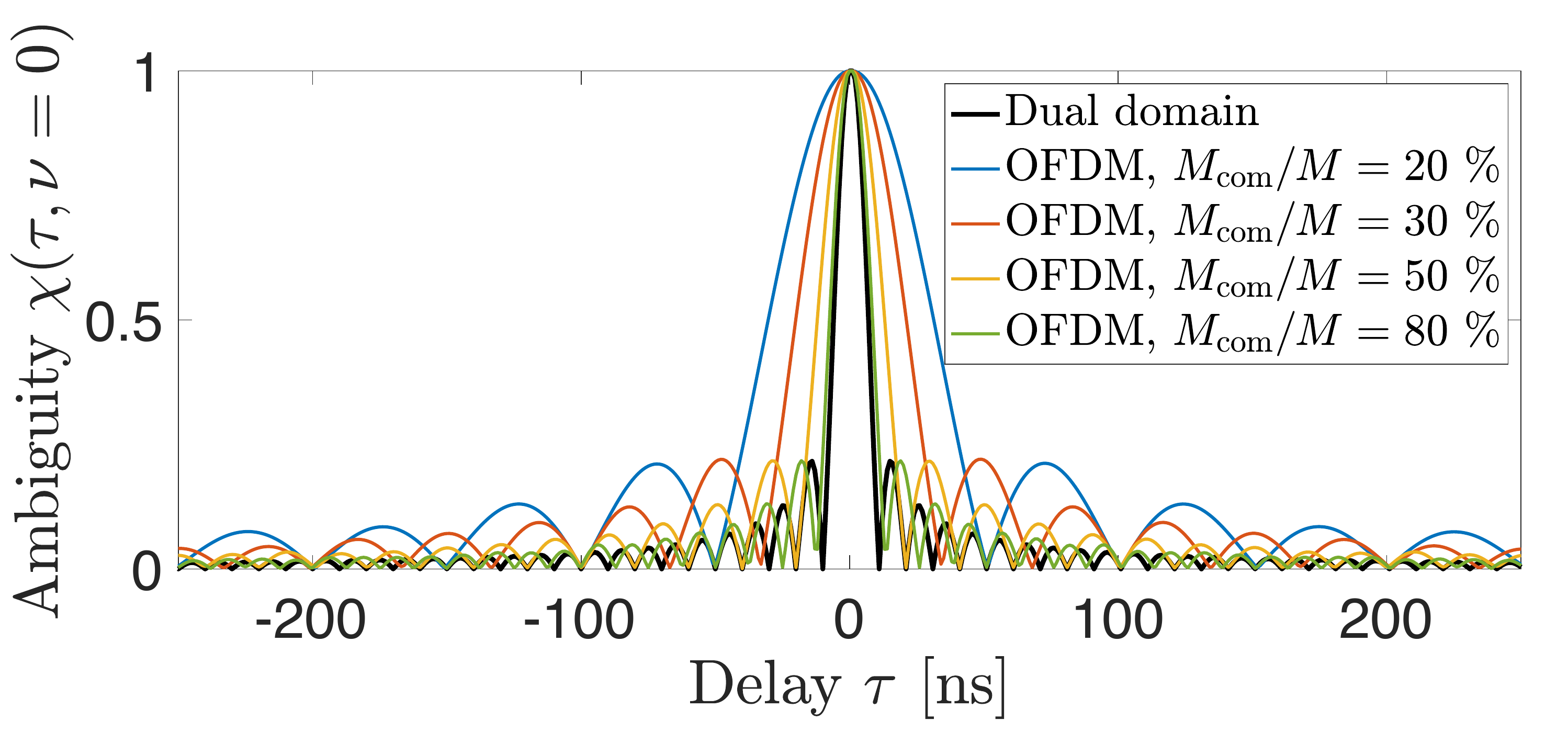}
    \caption{Slices of the ambiguity function of dual-domain and OFDM waveforms along the delay dimension, for different communication-allocated bandwidths and $\eta=100$ \% occupation}
    \label{fig:ambiguity}
\end{figure}

\begin{figure}[!t]
    \centering
    \includegraphics[width=0.6\columnwidth]{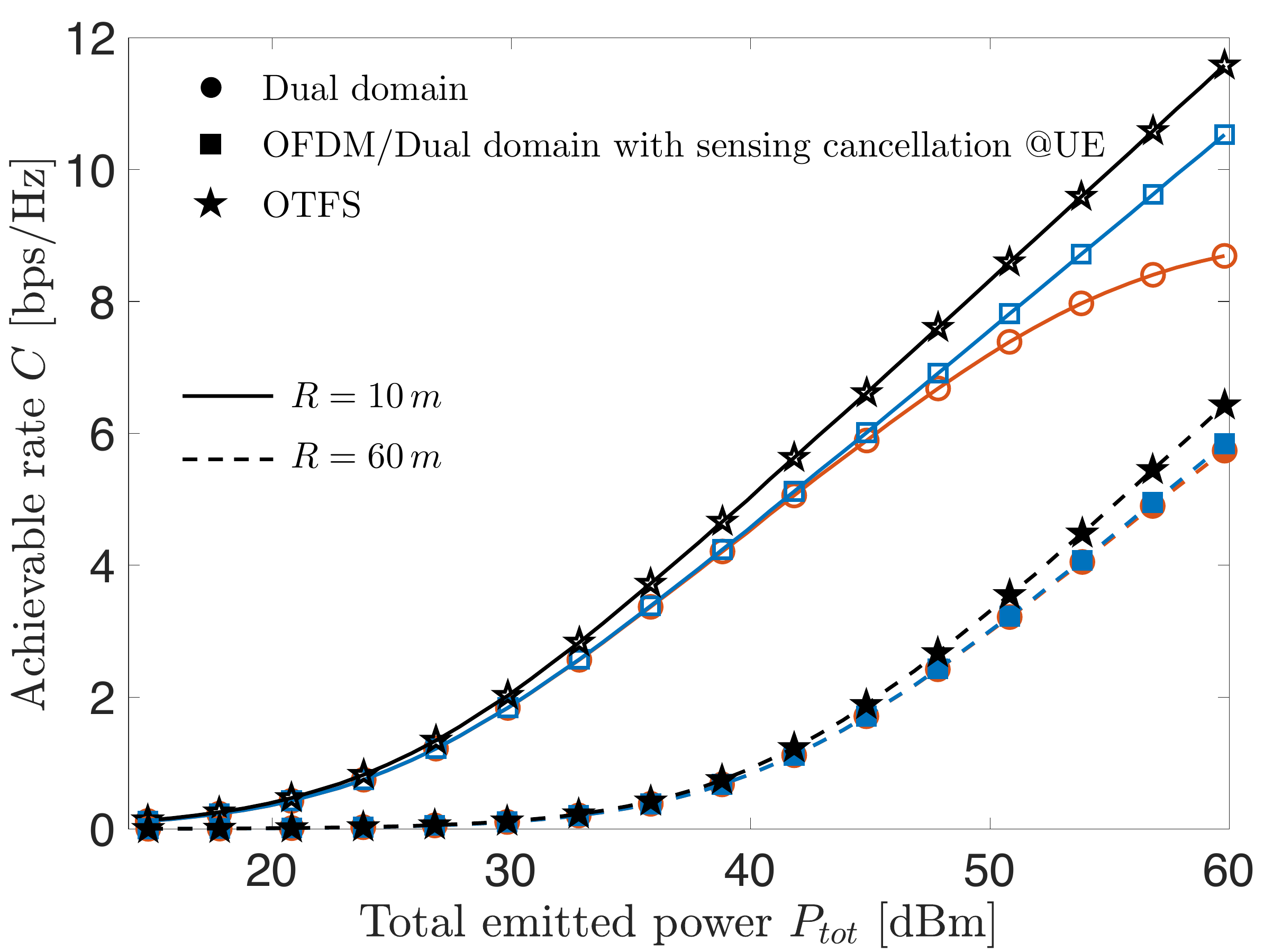}
    \caption{Achievable rate as a function of the total emitted power varying the BS-UE range, for dual-domain, OFDM and OTFS ISAC systems. Thanks to the proper power allocation for the sensing signal, there is no appreciable difference between the two systems, except for very high spectral efficiencies. }
    \label{fig:Rate}
\end{figure}

\begin{figure}[!t]
    \centering
%    \subfloat[][]{\includegraphics[width=1\columnwidth]{Figures/CRB_dual_OFDM_twotargets.pdf}\label{subfig:CRB_twotargets_absolute}}\\
%    \subfloat[][]{}
    \includegraphics[width=0.6\columnwidth]{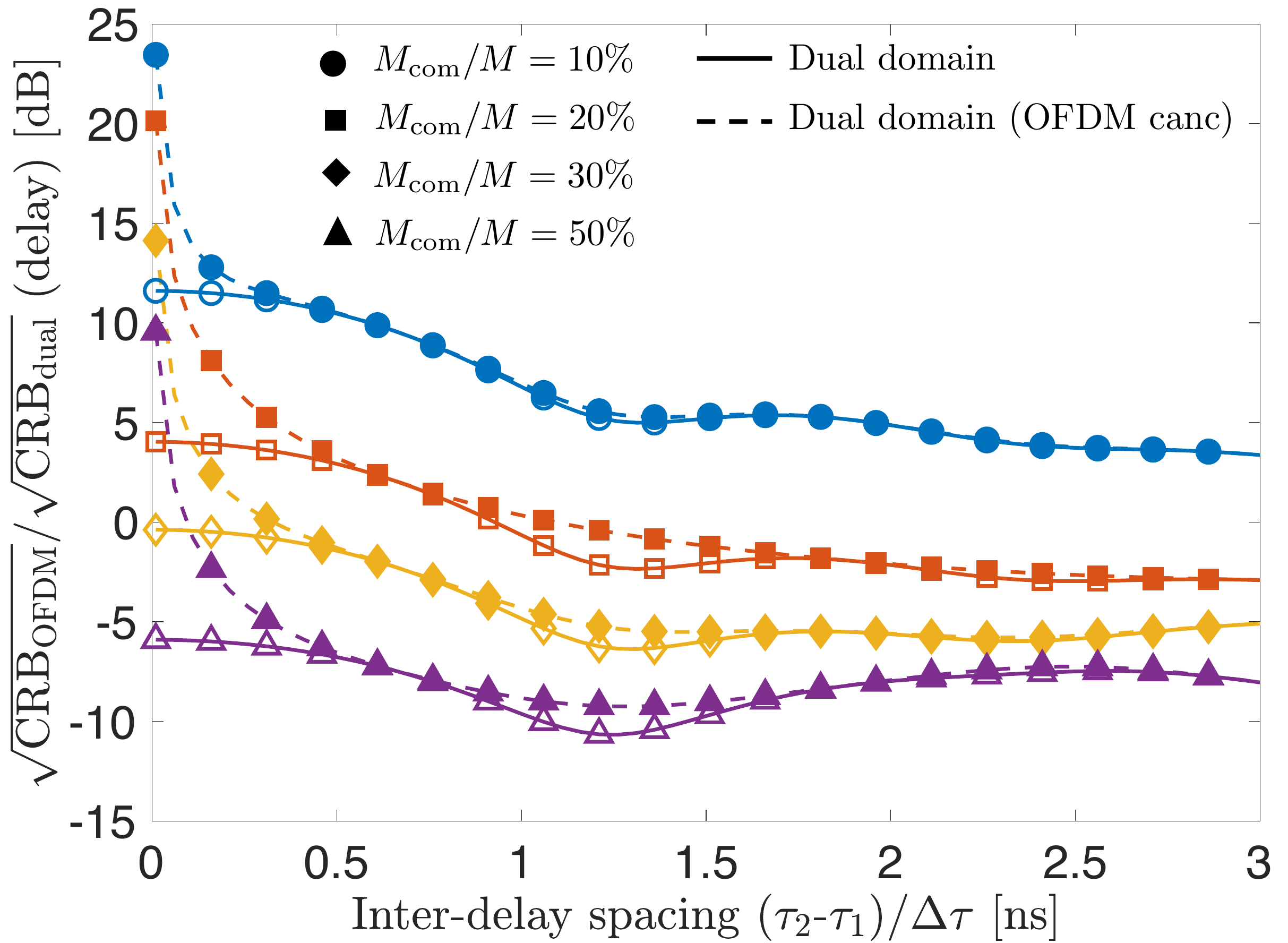}
    \caption{Root CRB ratio on delay estimation (two targets) between the OFDM and dual-domain waveforms, varying the normalized inter-delay spacing $(\tau_2-\tau_1)/\Delta \tau$.}
    \label{fig:CRB_twotargets_delay}
\end{figure}

%comaprison performance di sensing OFDM, OTFS e dual domain
The second set of results is related to the comparison between the sensing performance of the dual-domain, OFDM and OTFS waveforms, fixing the total emitted power to $P^\mathrm{FT}_{tot}=43$ dBm. As for Table \ref{tab:SimParam}, the per-subcarrier power of the sensing signal in the dual-domain ISAC waveform is $(\sigma^\mathrm{FT}_\mathrm{sen})^2 = 10^{-3}(\sigma^\mathrm{FT}_\mathrm{com})^2$ (30 dB less compared to communication one), thus it is 
\begin{equation}
    P_{\mathrm{com}}^\mathrm{FT} + P_{\mathrm{sen}}^\mathrm{FT} = P_{\mathrm{com}}^\mathrm{FT} + \frac{M}{M_\mathrm{com}}10^{-3}P_{\mathrm{com}}^\mathrm{FT} = P^\mathrm{FT}_{tot}
\end{equation}
The results highlight a power-bandwidth trade-off, suggesting that for $M_\mathrm{com}/M \leq 20\%$ (as currently available in 5G NR FR2 systems) the proposed dual-domain ISAC design is advantageous (comparable CRB, improved MTER) over OFDM and OTFS, notwithstanding a due power abatement for sensing. For higher OFDM/OTFS fractional bandwidths, these latter waveforms are preferable.
Figs. \ref{fig:Performance} and \ref{fig:ambiguity} summarize the results. 
Fig. \ref{subfig:CRB_delay} shows the ratio between the root CRB on delay estimation of the OFDM/OTFS and the root CRB of the dual-domain waveforms, as a function of the percentage of OFDM-occupied FT resources $\eta$ and for different fractional communication bandwidths $M_\mathrm{com}/M$. When the performance metric is $\geq 0$ dB we have a clear advantage in using dual-domain, otherwise OFDM/OTFS is better. As expected, for $M_\mathrm{com}/M > 30\%$, the superior Tx power employed by OFDM outperforms the larger bandwidth used by the sensing signal in the dual domain ISAC waveform, and the effect of the FT resource sparsity for OFDM ($\eta$) is mild. Notice that OTFS, by definition, has the same performance as OFDM for $\eta\rightarrow 1$ (full resource occupation).
It is however remarkable that, notwithstanding $(\sigma^\mathrm{FT}_\mathrm{sen}/\sigma^\mathrm{FT}_\mathrm{com})^2=-30$ dB, the dual-domain does not show any CRB penalty w.r.t OFDM/OTFS when $M_\mathrm{com}/M < 20$ \%. The usage of more bandwidth for sensing allows for filling the CRB gap with OFDM, thanks to a quadratic weighting of base-band frequencies (see Appendix \ref{app:CRB}), outside the communication spectrum. As an example, consider the case $M_\mathrm{com}/M = 10$ \%. The ratio between the effective bandwidth of the sensing signal and the OFDM one is $\propto (M/M_\mathrm{com})^3$, thus, perfectly compensating the 30 dB of power penalty.

However, the true advantage of using the dual-domain waveform is shown in Fig. \ref{subfig:MSLR}. This reports the MTER for the dual-domain, OFDM and OTFS. As the ambiguity function of the dual-domain sensing signal is a perfect 2D-sinc in DD domain, as shown in Fig. \ref{fig:ambiguity}, its MTER is around $80\%$. OTFS gets closer to this latter value but suffers a penalty due to the less Doppler resolution (see Appendix \ref{app:ambiguity}).
For all the OFDM waveforms, the MTER strongly reduces with $M_\mathrm{com}/M$. For instance, for $M_\mathrm{com}/M=30\%$, the MTER of OFDM drops from 20-to-40\%. This means that: \textit{(i)} the main lobe of the ambiguity function (along the delay) is $M_\mathrm{com}/M$-times larger (less resolution) and \textit{(ii)} some of the energy is spread over the whole DD domain (increased probability of masking weakly reflecting targets). The achieved ACLR, for the considered parameters, is no less than 18 dB, and can however be set according to regulatory limits by tuning the sensing power $P_\mathrm{sen}$. 

Finally, we discuss in Fig. \ref{fig:Rate} the achievable rate of the dual-domain, OFDM and OTFS systems, varying the Tx power and for $R=10,60$ m. The proposed dual-domain waveform is limited by the superposed 2D sensing sinusoid, but it results in a negligible penalty compared to OFDM, decreasing with the BS-UE distance, observed only for $C\geq 8$ bps/Hz. Of course, as the sensing signal is fixed and known, the UE can operate a suitable cancellation to reach the OFDM performance, at the price of a slightly increased complexity of the UE Rx. OTFS, instead, does not employ the CP, thus it gains over medium-to-high values of achievable rate.

\begin{figure}
    \centering
    \includegraphics[width=0.6\columnwidth]{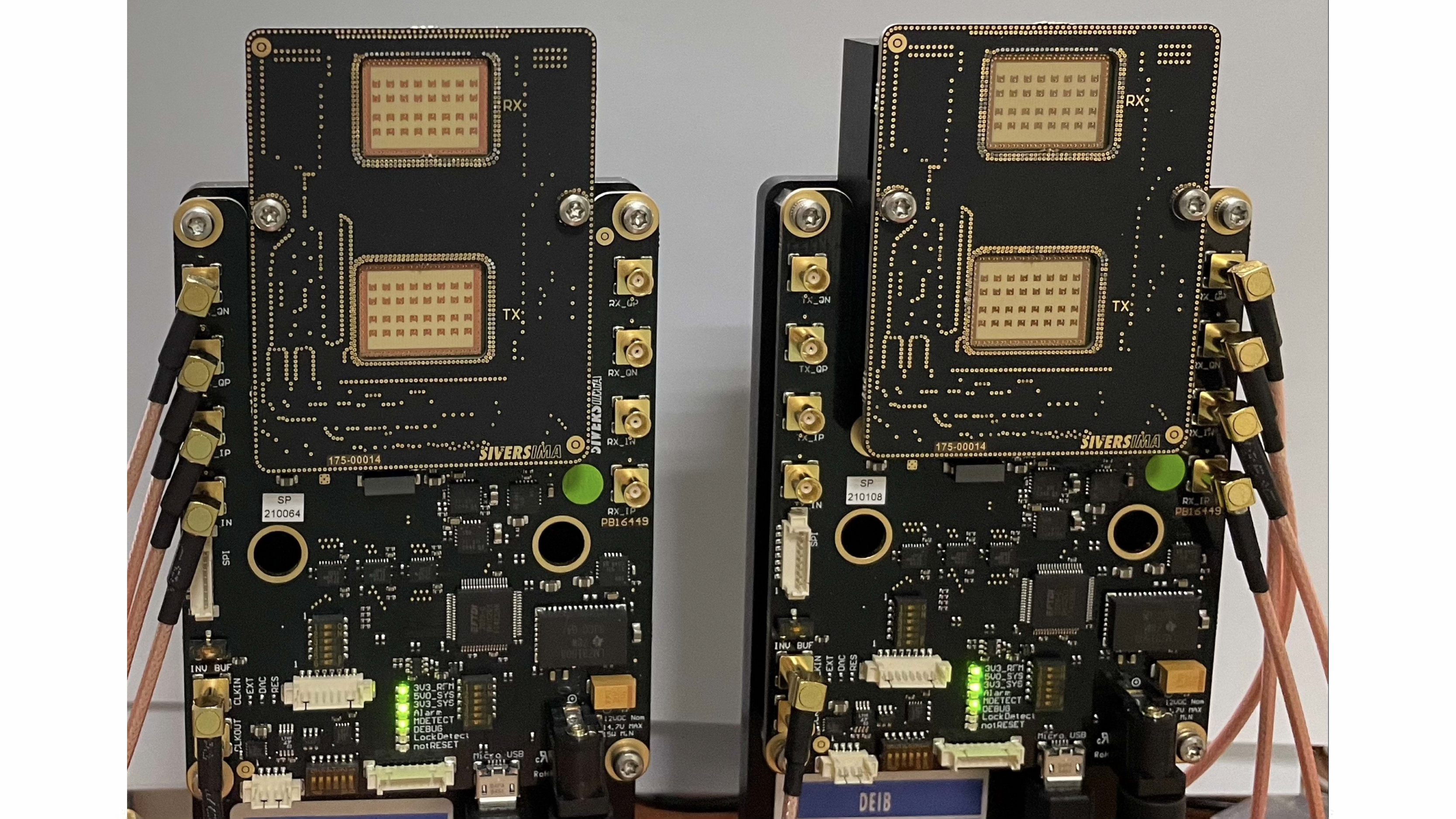}
    \caption{Sivers EVK06003 boards}
    \label{fig:setup}
\end{figure}

\begin{figure*}[!t]
    \subfloat[][Delay-Doppler Rx signal]{\includegraphics[width=0.45\columnwidth]{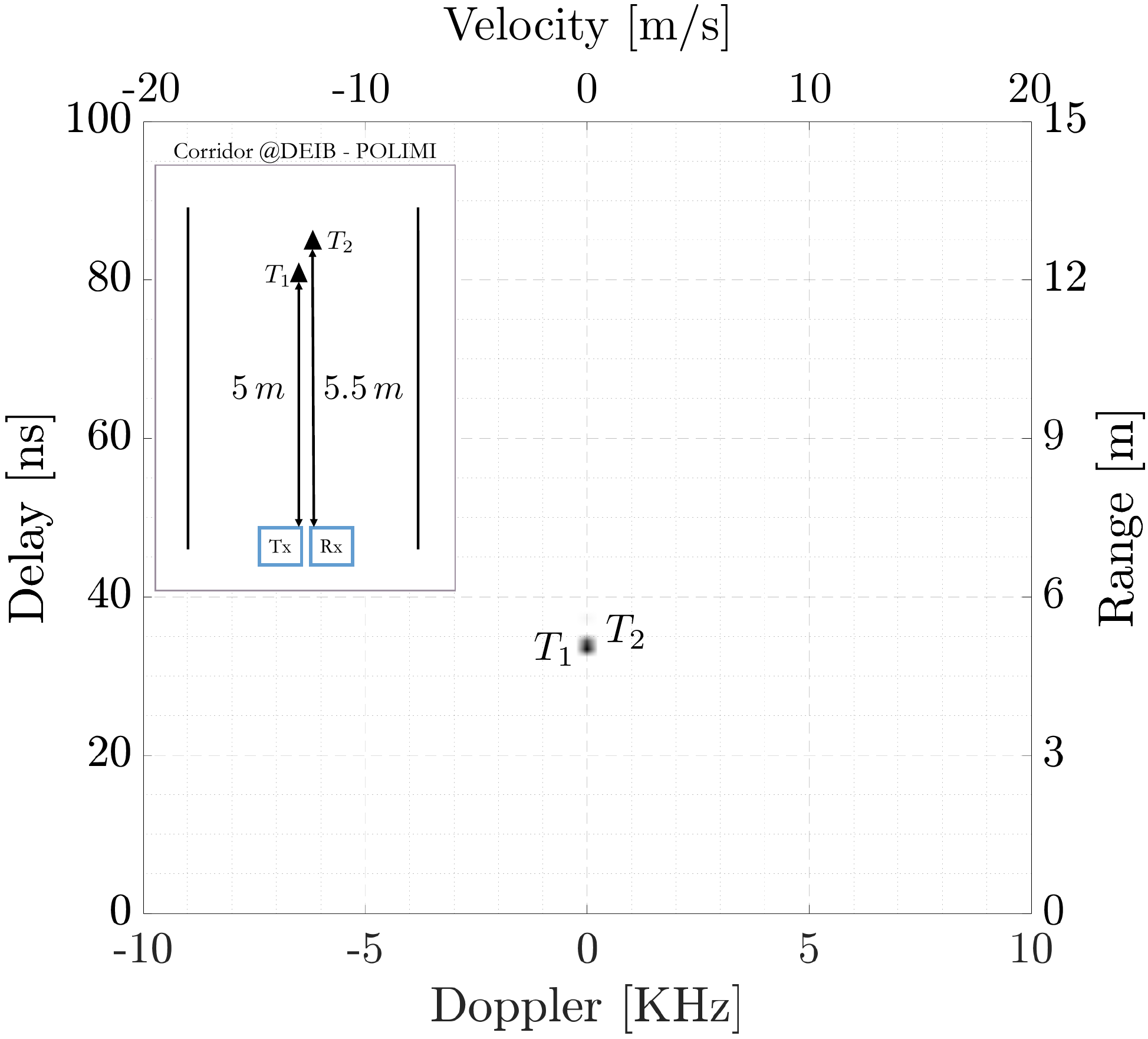}\label{subfig:static_DD}}
    \hspace{0.5cm}
    \subfloat[][Delay profile]{\includegraphics[width=0.45\columnwidth]{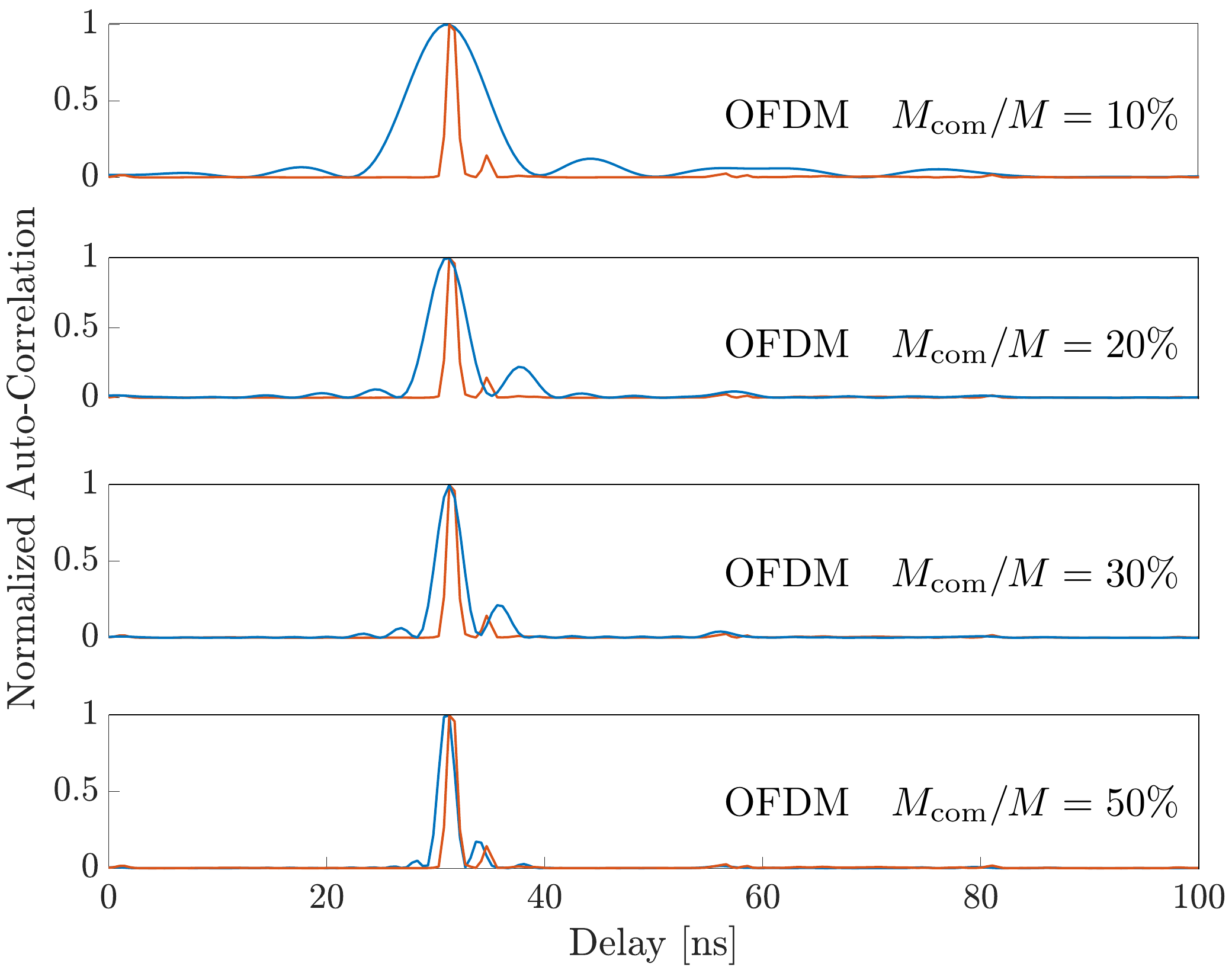}\label{subfig:static_profile}}\\
    \subfloat[][Rx constellation]{\includegraphics[width=0.6\columnwidth]{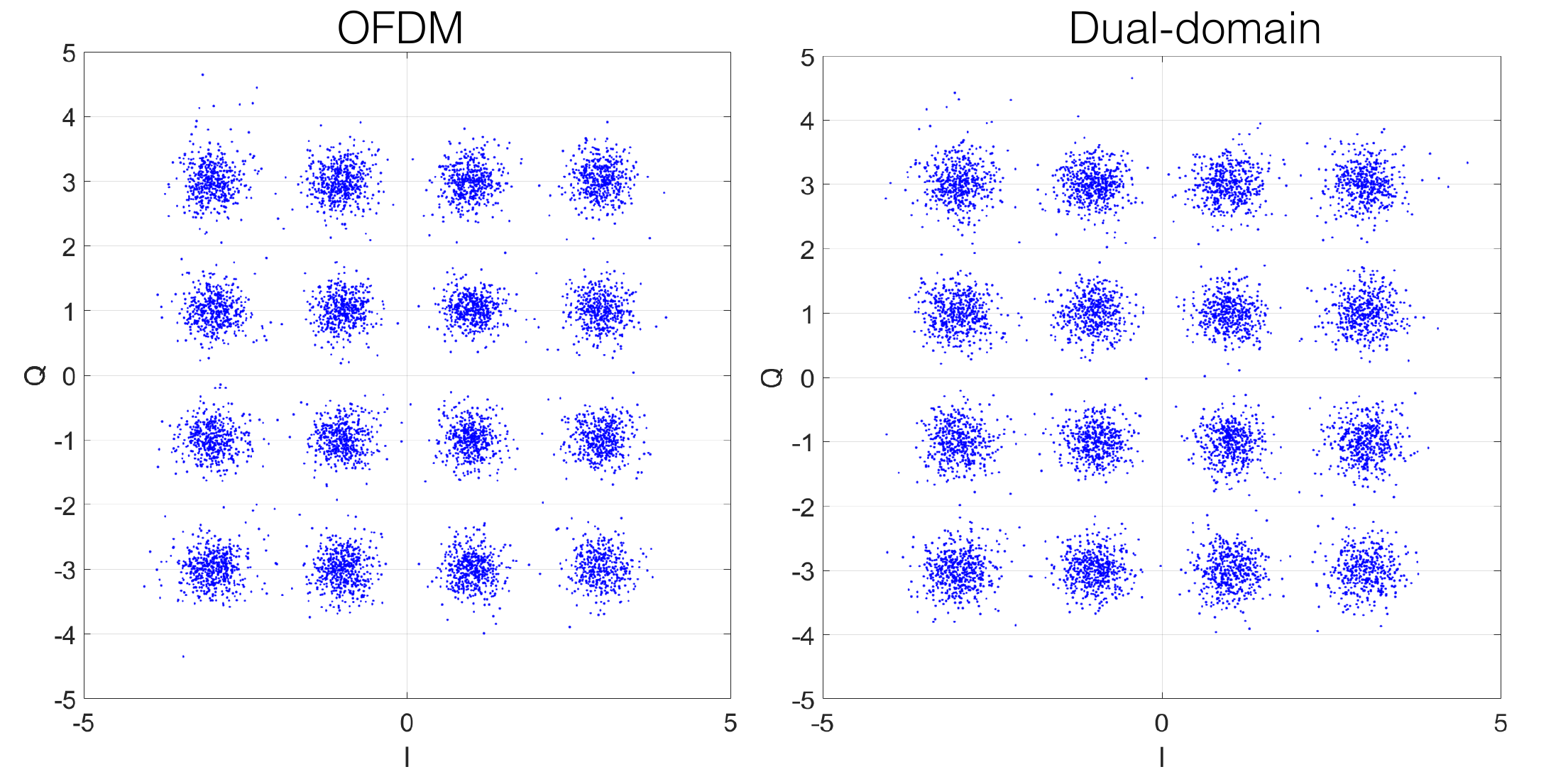}\label{subfig:static_constellation}}
    \caption{Experimental results for the two static targets scenario: (a) delay-Doppler Rx signal and (b) delay profile (c) Rx 16-QAM constellation with and without the sensing signal. }
    \label{fig:static_results}
\end{figure*}
\subsection{Sensing performance with two targets}

%two targets
The advantages of dual-domain ISAC waveform are even more evident when we consider the resolution of two closely-spaced targets along the delay/range dimension. Fig. \ref{fig:CRB_twotargets_delay} shows the root CRB ratio on delay estimation (two targets) between OFDM and dual-domain, varying the delay spacing $(\tau_2-\tau_1)/\Delta \tau$ (normalized to the maximum delay resolution $\Delta \tau = 1/(M\Delta f)$). The expression of the CRB is in Appendix \ref{app:CRB}, where we consider two equally strong targets ($\beta_2=\beta_1$). OTFS performance is herein equal to OFDM one and not shown. Again, values of the CRB ratio $\geq 0$ ($<0$) indicate an advantage in using dual-domain (OFDM), thus being representative of the \textit{gain} in using the proposed dual-domain ISAC waveform. 
Remarkably, the dual-domain allows resolving two closely spaced targets below the resolution limit ($\tau_2-\tau_1 \leq \Delta \tau$) with much higher accuracy compared to OFDM (root CRB gain in the order of 5-10 dB), thanks to a wider bandwidth (for $M_\mathrm{com}/M < 20$\%), irrespective of power abatement for the sensing signal. Compared to Fig. \ref{subfig:CRB_delay} (single target), the presence of two coupled targets pushes for the usage of more bandwidth. It is also interesting to notice that operating a cancellation of the OFDM signal at the BS side allows increasing the dual-domain performance of several dB in some cases.

%The root CRB gain of dual-domain over OFDM is shown in Fig. \ref{subfig:CRB_twotargets_ratio}, where we also report the ideal case where the OFDM communication signal in the dual-domain ISAC signal is perfectly cancelled at the BS. The CRB gain exceeds $10$ dB for $M_\mathrm{com}/M = 10$ \% (without OFDM cancellation) and $20$ dB (with OFDM cancellation). In any case, for the considered settings, the CRB penalty of dual-domain never exceeds 10 dB.

%capacità

%remark finale
The results underline the practical benefits of a dual-domain ISAC waveform, that allows increasing the delay/range resolution well beyond what is guaranteed by spectrum regulations (OFDM and OTFS) requiring no additional processing at the UE side.%, at the price of negligible penalties on the CRB and the communication rate.   

\section{Experimental Demonstration}\label{sect:experimental_results}

\begin{figure}[!t]
    %\subfloat[][Setup]{\includegraphics[width=\columnwidth]{Figures/Setup.png}\label{subfig:Setup}}\\  
    %\caption{Experimental setup employed for the demonstration, with (\ref{subfig:Setup}) ISAC transceiver . The HW equipment consists of a high-resolution imaging radar, used as a benchmark, and two communication transceivers, one used as Tx and the other as Rx. }
    %\label{fig:experimental_setup}
    \includegraphics[width=0.6\columnwidth]{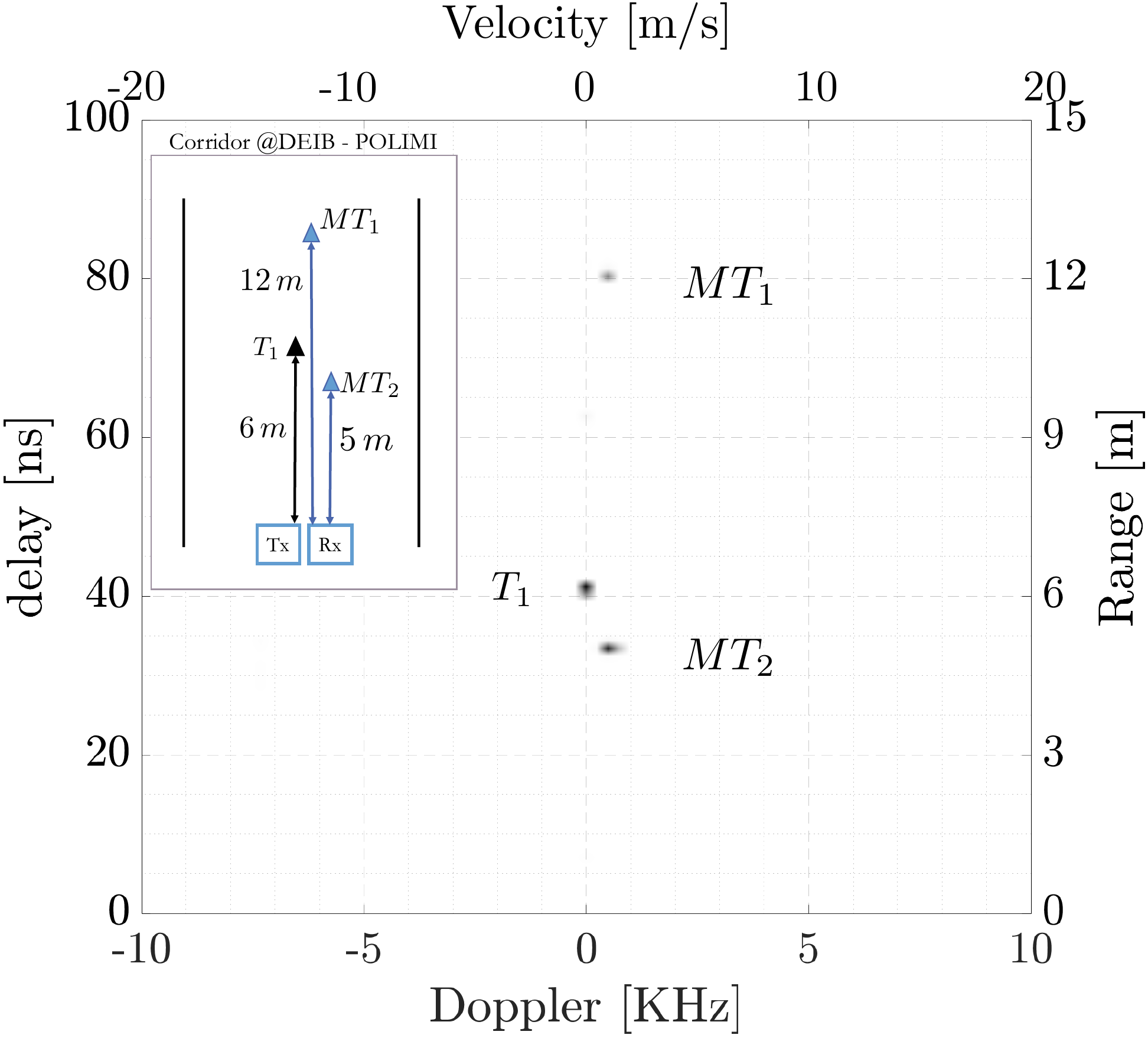}
    \caption{Delay-Doppler Rx signal for one static and two moving targets, the latter two in the same direction with slightly different velocities. }\label{fig:dynamic_results}
\end{figure}

To further demonstrate the feasibility of the proposed ISAC waveform, we experimentally test its performance. The setup is composed of a pair (one Tx and one Rx) of mm-Wave EVK06003 communication transceivers from Sivers semiconductors, operating at $f_0=60$ GHz, over a bandwidth of $B=1$ GHz and featuring analog $8\times 2$ arrays \cite{Sivers} (Fig. \ref{fig:setup}). The two EVK06003 transceivers operate as Tx for ISAC and Rx for sensing, and one single-antenna Rx is used as a communication Rx. A Xilinx ZCU111 FPGA board controls all the RF boards \cite{Xilinx}, which are time and frequency synchronized. 

The first experiment includes the range estimation of two closely-spaced static metallic targets (corner reflectors), respectively placed at $5$ and $5.5$ m from the Tx/Rx, and the reception of the communication signal by a terminal at $4$ m.
The Tx dual domain ISAC signal generation follows the procedure in Section \ref{sect:TxISAC}, employing the available Tx power $P_{tot} = 0$ dBm, fractional bandwidth for communication $M_\mathrm{com}/M=10,20,30,50$\%, subcarrier spacing $\Delta f=1$ MHz, number of OFDM symbols $N=4096$ (with duration $T=1$ $\mu$s, thus using a burst of approx. 4 ms), OFDM in-band resource occupation $\eta=1$ and $(\sigma^\mathrm{FT}_s/\sigma^\mathrm{FT}_\mathrm{com})^2 = 10^{-4}$ ($-40$ dB) to ensure that the Tx ISAC signal does not saturate the full-scale range of the digital-to-analog converter and such that the Tx impulse train corresponding to the 2D sensing sinusoid has the same peak amplitude of the OFDM signal. Figure \ref{fig:static_results} shows both the DD Rx signal (Fig. \ref{subfig:static_DD}) and the delay profile (normalized autocorrelation) of the Rx signal (for zero Doppler, Fig. \ref{subfig:static_DD}). The former is obtained with the proposed ISAC signal, while the latter compares the delay profile of the proposed ISAC signal (red curves) with the one achieved with OFDM (blue curves), varying $M_\mathrm{com}/M$. Remarkably, the proposed ISAC signal allows clearly distinguishing the two targets in the delay/range domain, notwithstanding the power abatement to cater ACLR regulations (the achieved relative ACLR is $\mathrm{ACLR}_\mathrm{rel}= 28.8$ dB. Differently, OFDM suffers from bandwidth limitations and, for $M_\mathrm{com}/M \leq 30 \%$, provides the range of the first (and strongest) target only. As predicted by numerical results in the previous section, Fig. \ref{subfig:static_constellation} shows that the presence of the sensing signal does not affect the communication, herein a 16-QAM is considered, as the SNR at the communication Rx is 18.71 dB (with only OFDM) while the SDNR is 18.18 dB (for dual-domain). 
As a further confirmation of the validity of the proposed ISAC waveform, we also report in Fig. \ref{fig:dynamic_results} the DD Rx signal for a static target at $6$ m distance from Tx/Rx and two moving targets at $5$ m and $12$ m respectively, characterized by the same radial velocity of $1.22$ m/s (positive for a motion in the direction of Tx/Rx). Again, using a sensing signal with a very limited emitted power but a large spectrum and sufficient duration, we can detect both static and moving targets.    
These results demonstrate the practical feasibility of the proposed dual-domain ISAC waveform to increase the resolution of bandwidth-constrained communication systems.  
%The resulting range-angle images are shown in Fig. \ref{fig:images}. The TI board provides superior resolution performance and allows clearly identifying the two targets, located in the correct position within the range-angle resolution of the system, and the surrounding clutter. The latter mostly comprises metal gratings and concrete columns (making strong dihedral reflectors with the walls). The image generated by the Sivers EVK06003 setup is less rich in detail, some details are missing, but the two reference targets are however visible. It has to be noticed that, while the range of the two is correctly estimated ($9.06$ m and $6.1$ m), the angle is distorted due to slight angular placement errors of the Sivers boards. This issue does not affect the quality of the proposed ISAC-generated image in practical systems, where Tx and Rx are manufactured on the very same board. This result demonstrates the practical feasibility of the proposed dual-domain ISAC waveform to increase the resolution of bandwidth-constrained communication systems.  

%
\section{Conclusion}\label{sect:conclusion}
This paper proposes a novel dual-domain ISAC waveform design for communication-centric approaches, based on the superposition of a properly designed large bandwidth sensing signal with the legacy communication one (OFDM) in the FT domain. Through the definition of a proper power allocation problem for the sensing signal, the latter can coexist with the communication one without interference at the UE side. A peculiarity of the proposed approach is that it allows the ISAC system to exceed regulatory spectrum limits with the aim of enhancing delay/range resolution in medium-to-short ranges (60-100 m). Numerical results show the benefits of the proposed ISAC method by comparing it with OFDM and OTFS in terms of CRB, ambiguity function and achievable rate. In particular, although sensing signal power is reduced by 30 dB compared to the communication one to comply with out-of-band emission limitations, the CRB on delay estimation of the dual-domain ISAC waveform is comparable with the OFDM/OTFS one  when the excess bandwidth is around 80-90\%. Remarkably, the achievable rate is not practically affected by the presence of the sensing signal. An experimental demonstration shows the feasibility of the proposed method, suggesting its usage as a possible solution for high-resolution ISAC systems. 

%In this paper, we propose a novel JC\&S waveform to enable sensing capabilities on top of a communication system, based on the superposition of a properly designed signal in the DD domain and an OFDM communication signal in the FT domain. The dual domain JC\&S waveform is first analytically described, pointing out the relations of the signals in different domains and their mutual interference. The proposed waveform is able to carry information to multiple UEs while sensing the environment, estimating range and velocity of the UEs.
%Numerical results show that, with a fixed power budget and a proper power allocation to communication and sensing, the capabilities of the latter (both resolution) can be tuned by the choice of the available bandwidth and length of the downlink burst, while leaving the BER substantially unaffected. 

%
\appendices

\section{}\label{app:ambiguity}

The ambiguity function of a signal for the $K$ UEs is dependent on the specific resource allocation strategy. In case of $g(t)=\mathrm{rect}(t/\Delta\tau)$, we have \cite{Wymeersch2021}:
\begin{equation}\label{eq:ambiguity}
\begin{split}
    \chi(\tau,\nu) & \simeq \mathbf{p}^\mathrm{T} \left(\boldsymbol{\nu}^*(\nu) \otimes \boldsymbol{\tau}(\tau)\right)
\end{split}
\end{equation}
for $-1/(2T)\leq\nu\leq 1/(2T)$ and $0\leq \tau \leq T'$, where $\mathbf{p}=\left(\mathbf{s}\odot\mathbf{s}^*\right)\in\mathbb{R}^{MN\times 1}$ is the vector of power allocation across FT resources ($\mathbf{s}\in\mathbb{C}^{MN\times 1}$ is the Tx signal), and   while 
\begin{align}
    \boldsymbol{\tau}(\tau) &= [e^{-j2\pi m \Delta f \tau}]_{m=-\frac{M}{2}}^{\frac{M}{2}-1}\label{eq:delay_response_vector}\\
    \boldsymbol{\nu}(\nu) &= [e^{j2\pi n T \nu}]_{n=-\frac{N}{2}}^{\frac{N}{2}-1}\label{eq:Doppler_response_vector}
\end{align}
are the frequency and time channel responses to a delay $\tau$ and a Doppler shift $\nu$. 
For OFDM signal, we have $\mathbf{p}=\mathbf{p}_\mathrm{com} = \mathrm{vec}\left(\boldsymbol{\Sigma}^\mathrm{FT}_\mathrm{com} \odot \boldsymbol{\Sigma}^\mathrm{FT}_\mathrm{com}\right)$. For the proposed dual-domain waveform, it is $\mathbf{p}=\left(\sigma^{\mathrm{FT}}_\mathrm{sen}\right)^2 \mathbf{1}^\mathrm{T}_{MN}$, as the ambiguity is made by the sensing signal only. For OTFS, instead, the Tx signal in FT domain occupies all the resources, such as in the dual-domain case, but there is no CP in the Tx waveform. Thus, the ambiguity is approximated as in \eqref{eq:ambiguity} where $\boldsymbol{\nu}(\nu)$ is defined with $T'$ instead of $T$. This means that, for fixed number of symbols $N$, the OTFS has less Doppler resolution compared to OFDM and dual-domain but the achievable rate is higher. 

%The ambiguity function of the dual domain ISAC signal is well approximated, after the matched filter (IDFT+DFT), by the ambiguity function of the sensing signal only:
%\begin{equation}
%\begin{split}
%    \chi_{\text{dual}}(\tau,\nu) &  \simeq \left(\sigma^{\mathrm{FT}}_s\right)^2 \mathbf{1}^\mathrm{T}_{MN} \left(\boldsymbol{\nu}^*(\nu) \otimes \boldsymbol{\tau}(\tau)\right).
%\end{split}
%\end{equation}
%
The energy of the main lobe of the ambiguity function is
\begin{equation}\label{eq:energy_mainlobe}
E_{\text{main}}=\sqrt{\int\limits_{-\frac{1}{2M\Delta f}}^{\frac{1}{2M\Delta f}}
    \int\limits_{-\frac{1}{2NT}}^{\frac{1}{2NT}}  \lvert\chi(\tau,\nu)\rvert^2\, d\nu d\tau}, 
\end{equation}
which is evaluated considering the total available bandwidth $B=M\Delta f$. The MTER is 
\begin{equation}
    \mathrm{MTER} = \frac{E_{\text{main}}}{E}.
\end{equation}
where $E$ is the total energy of the waveform.

\section{}\label{app:CRB}
We derive the CRB for a generic signal in the FT domain defined over the resource grid $\Lambda^{\mathrm{FT}}$, for two targets whose delays are $\tau_1$ and $\tau_2$, Doppler shifts $\nu_1$ and $\nu_2$ and scattering amplitudes $\beta_1$ and $\beta_2$. The following derivation is valid for the OFDM and the dual-domain ISAC waveforms. The Rx signal in the FT domain is
\begin{equation}
    \mathbf{y} = \beta_1 \, \mathbf{s} \odot \boldsymbol{\tau}(\tau_1) \otimes \boldsymbol{\nu}(\nu_1) + \beta_2 \, \mathbf{s} \odot \boldsymbol{\tau}(\tau_2) \otimes \boldsymbol{\nu}(\nu_2) + \mathbf{z}
\end{equation}
where $\mathbf{y}\in\mathbb{C}^{MN \times 1}$ is the vector of Rx signal in FT domain, $\mathbf{s}$ is the vector of Tx signal, $\mathbf{z}\sim\mathcal{CN}(\mathbf{0},\mathbf{R}_z)$ is the additive Gaussian noise.

Consider for simplicity the upper-bound in which delay and Doppler estimation are decoupled, thus the Fisher information matrix (FIM) is $\mathbf{I} = \mathrm{blkdiag}( \mathbf{I}_{\boldsymbol{\tau}}, \mathbf{I}_{\boldsymbol{\nu}})$, where $\mathbf{I}_{\boldsymbol{\tau}}\in\mathbb{R}^{2\times 2}$ is related to delay and $\mathbf{I}_{\boldsymbol{\nu}}\in\mathbb{R}^{2\times 2}$ is related to Doppler. We have the following Fisher information terms: 
\begin{equation}\label{eq:CRB_tau1tau1}
    I_{\tau_1,\tau_1} = 2 |\beta_1|^2\mathbf{p}^\mathrm{T}\left( \mathrm{diag}\left(\mathbf{R}^{-1}_z\right)\odot 4\pi^2 \Delta f^2 (\mathbf{m}\hspace{-0.05cm}\odot\hspace{-0.05cm}\mathbf{m}) \otimes \mathbf{1}_N\right)
\end{equation}
\begin{equation}\label{eq:CRB_tau2tau2}
    I_{\tau_2,\tau_2} = 2 |\beta_2|^2\mathbf{p}^\mathrm{T}\left( \mathrm{diag}\left(\mathbf{R}^{-1}_z\right)\odot 4\pi^2 \Delta f^2 (\mathbf{m}\hspace{-0.05cm}\odot\hspace{-0.05cm}\mathbf{m}) \otimes \mathbf{1}_N\right)
\end{equation}
\begin{equation}\label{eq:CRB_tau1tau2}
    I_{\tau_1,\tau_2} = 2\, \mathbf{p}^\mathrm{T}\Re\left\{ \mathrm{diag}\left(\mathbf{R}^{-1}_z\right)\odot 4\pi^2 \beta_1^*\beta_2 \Delta f^2 (\mathbf{m}\hspace{-0.05cm}\odot\hspace{-0.05cm}\mathbf{m}\odot \boldsymbol{\tau}_{1,2} ) \otimes \boldsymbol{\nu}_{1,2}\right\}
\end{equation}
\begin{equation}\label{eq:CRB_nu1nu1}
    I_{\nu_1,\nu_1} = 2 |\beta_1|^2 \mathbf{p}^\mathrm{T} \left(\mathrm{diag}\left(\mathbf{R}^{-1}_z\right)\odot (\mathbf{1}_M \otimes 4\pi^2 T^2 (\mathbf{n}\odot\mathbf{n})) \right)
\end{equation}
\begin{equation}\label{eq:CRB_nu2nu2}
    I_{\nu_2,\nu_2} = 2 |\beta_2|^2 \mathbf{p}^\mathrm{T} \left(\mathrm{diag}\left(\mathbf{R}^{-1}_z\right)\odot (\mathbf{1}_M \otimes 4\pi^2 T^2 (\mathbf{n}\odot\mathbf{n})) \right)
\end{equation}
\begin{equation}\label{eq:CRB_nu1nu2}
    I_{\nu_1,\nu_2} = 2 \,\mathbf{p}^\mathrm{T} \Re\left\{\mathrm{diag}\left(\mathbf{R}^{-1}_z\right)\odot (\boldsymbol{\tau}_{1,2} \otimes 4\pi^2 T^2 \beta_1^*\beta_2(\mathbf{n}\odot\mathbf{n})\odot \boldsymbol{\nu}_{1,2}) \right\}
\end{equation}
where $\mathbf{n}=[-(N/2),...,(N/2)-1]$, $\mathbf{m}=[-(M/2),...,(M/2)-1]$ and
\begin{equation}
\begin{split}
    \boldsymbol{\tau}_{1,2} = \mathrm{diag}(\boldsymbol{\tau}(\tau_1)\boldsymbol{\tau}(\tau_2)^\mathrm{H}),\\
    \boldsymbol{\nu}_{1,2} = \mathrm{diag}(\boldsymbol{\nu}(\nu_1)\boldsymbol{\nu}(\nu_2)^\mathrm{H})
\end{split}
\end{equation}
are the cross-coupled delay and Doppler channel responses.
For OFDM-based ISAC, only a subset of the time-frequency resources are allocated, and $\mathbf{R}_z = \sigma^2_z\mathbf{I}_{MN}$. Differently, for the dual-domain waveform, all the time-frequency resources are occupied by the sensing signal, and the communication one acts as interference. Thus, we have $\mathbf{p} = (\sigma^\mathrm{FT}_s)^2\mathbf{1}_{MN}$ and we can model the noise covariance matrix as
\begin{equation}
    \mathbf{R}_z = \sigma^2_z\mathbf{I}_{MN} + (|\beta_1|^2 + |\beta_2|^2) \mathbf{p}_\mathrm{com}
\end{equation}
where is the vector of power allocated on each single FT resource unit (subcarrier-symbol) for communication.
The CRB on delay and Doppler is therefore lower-bounded by
\begin{align}
    \mathbf{CRB}_\tau  = \mathbf{I}_{\boldsymbol{\tau}}^{-1},\qquad \mathbf{CRB}_\nu = \mathbf{I}_{\boldsymbol{\nu}}^{-1}.
\end{align}

\section*{Acknowledgment}
The work is framed within the Huawei-Politecnico di Milano Joint Research Lab. 

\bibliographystyle{IEEEtran}
\bibliography{Bibliography}

\end{document}